\documentclass[aps,prl,epsfigure,twocolumn,superscriptaddress]{revtex4-2}
\usepackage[colorlinks=true,linkcolor=blue,urlcolor=blue,citecolor=blue,pdfusetitle]{hyperref}
\usepackage[utf8]{inputenc}
\usepackage[english]{babel}
\usepackage{amsmath}
\usepackage[caption = false]{subfig}
\usepackage{graphicx,epstopdf}
\usepackage{blindtext}
\usepackage[table,xcdraw]{xcolor}
\usepackage{lipsum}
\usepackage{amsfonts}
\usepackage{bbm}
\usepackage{amssymb}
\usepackage{enumerate}
\usepackage{color}
\usepackage{latexsym}
\usepackage{physics}
\usepackage{times,txfonts}

\newcommand{\Dcal}{\mathcal{D}}

\newcommand{\Fcal}{\mathcal{F}}

\newcommand{\Pcal}{\mathcal{P}}

\newcommand{\1}{\mathbbm{1}}

\newcommand{\SubFig}[2]{\ref{#1}{\color{blue}#2}}

\definecolor{bluePoli}{cmyk}{0.4,0.1,0,0.4}
\definecolor{blueGreen}{RGB}{44, 120, 247}
\definecolor{brickred}{rgb}{0, 0, 0}
\definecolor{darkred}{RGB}{204, 0, 0}
\definecolor{darkgreen}{RGB}{0, 102, 50}
\definecolor{darkblue}{RGB}{0, 76, 153}
\definecolor{mygold}{RGB}{255, 128, 32}
\definecolor{mypurple}{RGB}{178, 105, 252}
\definecolor{myorange}{RGB}{204, 102, 0}

\newcommand{\RefA}[1]{{\color{brickred}#1}}

\newcommand{\CSIC}{Instituto de Física Fundamental, Consejo Superior de Investigaciones Científicas, Calle Serrano 113b, 28006 Madrid, Spain}

\newcommand{\gdpkl}{Southern University of Science and Technology, Shenzhen, Guangdong 518055, China}
\newcommand{\szkl}{International Quantum Academy, Futian District, Shenzhen, Guangdong 518048, China}
\newcommand{\HFNL}{Shenzhen Branch, Hefei National Laboratory, Shenzhen 518048, China}

\usepackage{orcidlink}

\begin{document}

\title{Collective Dynamics in Circuit Quantum Acoustodynamics with a Macroscopic Resonator} 
\author{Libo Zhang}
\thanks{These authors contributed equally to this work.}
\affiliation{\szkl}\affiliation{\gdpkl}

\author{Chilong Liu}
\thanks{These authors contributed equally to this work.}
\affiliation{\szkl}\affiliation{\gdpkl}

\author{Guixu Xie}
\thanks{These authors contributed equally to this work.}
\affiliation{\szkl}\affiliation{\gdpkl}

\author{Haolan Yuan}
\thanks{These authors contributed equally to this work.}
\affiliation{\szkl}\affiliation{\gdpkl}

\author{Mingze Liu}
\affiliation{\szkl}\affiliation{\gdpkl}

\author{Hao Jia}
\affiliation{\szkl}

\author{Jian Li}
\affiliation{\szkl}

\author{Chang-Kang Hu}
\email{huchangkang@iqasz.cn}
\affiliation{\szkl}

\author{Song Liu}
\email{lius@iqasz.cn}
\affiliation{\szkl}\affiliation{\HFNL}

\author{Alan C. Santos~\orcidlink{0000-0002-6989-7958}}
\email{ac\_santos@iff.csic.es}
\affiliation{\CSIC}

\author{Dian Tan}
\email{tandian@iqasz.cn}
\affiliation{\szkl}\affiliation{\HFNL}

\author{Dapeng Yu}
\affiliation{\szkl}\affiliation{\HFNL}


\begin{abstract}
Collective dynamics in engineered quantum systems offer a unique and versatile platform for exploring how many-body correlations bridge microscopic entanglement and macroscopic behavior. In this work, we report \RefA{collective Dicke dynamics} of acoustic modes in a macroscopic high-overtone bulk acoustic resonator (HBAR). To achieve this, we engineer a hybrid quantum acoustodynamic system comprising an HBAR strongly coupled to a superconducting transmon qubit. The HBAR device is distinctive in the sense that its narrow mode spacing, together with enhanced qubit-mode coupling strength, gives rise to efficient coupling between the transmon and clusters of near-resonant modes.
By harnessing the system properties, we observe collective dynamics involving clusters composed by two or three mechanical modes, where their non-resonant spectrum allows for the observation of the transition between the Dicke \textit{static} regime to dynamically induced \textit{timed}-Dicke one. The coherent collective behavior of the system is supported by time-domain measurements of  the qubit's purity, \RefA{indicating} the quantum nature of the collective dynamics. Overall, our work establishes HBAR-based hybrid quantum system as a promising platform for exploring many-body collective dynamics in macroscopic mechanical systems.
\end{abstract}

\maketitle

\textit{Introduction.---} 
An intriguing phenomena in many-body interacting systems is the emergence of collective behavior, where multiple quantum elements interact strongly, resulting in novel dynamics that are central to both fundamental quantum mechanics~\cite{Dick1954,Gross:82} and emerging quantum technologies~\cite{Haake:93,Bohnet:12,Scully:15,Perarnau:20,Sheremet:23}. In classical systems, collective dynamics arise naturally due to mutual interactions, but in quantum systems, these phenomena are far richer due to the underlying quantum correlations and coherence~\cite{Ritsch2013, Mahdi2004, Finn2022, Tao2025}. Extensive experimental studies with atoms have demonstrated the existence of collective quantum phenomena~\cite{Anderson1995,Fink2009,Wang:20}. 
Yet, such effects are less studied in bosonic networks, as nonlinear collective behavior are not expected and nonlinearities in such a system are experimentally complex with strong interactions. 

In recent years, the field of quantum acoustics has witnessed remarkable progress in creating hybrid quantum systems ~\cite{Hybrid20,Chu2020APL} that seamlessly link superconducting qubits with mechanical resonators, opening new avenues for quantum information processing and fundamental physics at \RefA{single-excitation level}~\cite{Connell10, Teufel11, Martin14, Manenti:17, Chu:17, Satzinger18, Andersson19, Wollack2022,arrango2019,Qiao2023}. Notable achievements include quantum control of single phonons~\cite{Connell10, Chu:17, Satzinger18,Chu:18,Yiwen2023,Yiwen2024}, the creation of entangled phonon states~\cite{Wollack2022, Cleland1, Cleland2,von_Lupke:24}, among others~\cite{Kervinen19, Mirhosseini20,Yohannes2023, Bozkurt25, Lin2025, arXiv1, arXiv2}. In this scenario, HBARs have emerged as a promising platform for studying multi-mode dynamics in macroscopic quantum systems due to its high quality factor and efficient quantum control techniques enabling strong coupling between the phonon modes with a superconducting qubit~\cite{Epitaxy20, Chu:18}.

\begin{figure*}[t!]
	\includegraphics[width=\linewidth]{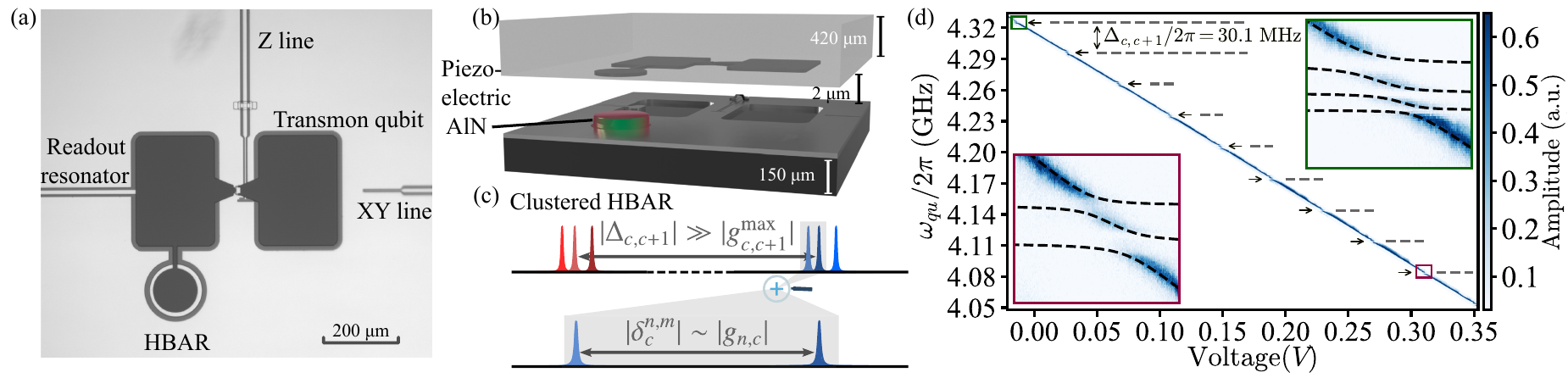}
	\caption{The hybrid quantum system. (a) Microscope image of the experimental device, featuring a superconducting transmon qubit coupled to a high-overtone bulk acoustic resonator (HBAR). 
		(b) Schematic of the device, illustrating the hybrid quantum system fabricated by a flip-chip technique to align the bottom chip with the top chip. Top: the flux-tunable qubit with an antenna for coupling the HBAR; Bottom: The qubit control lines, The HBAR,  readout resonators and a Purcell filter. 
		(c) Representation of clustered modes obtained in our device, where the top figure shows two clustered modes separated in frequency by $\Delta_{c,c+1}$, much greater than the qubit-HBAR coupling strength. By analyzing the structure of each cluster, bottom figure, it shows that the intra-mode spacing $\delta^{n,m}_{c}$ is comparable to the coupling strength. 
		(d) Spectroscopy of the qubit $Q_B$, experimentally showing nine different clusters with inter-mode FSR around $\Delta_{c,c+1} \approx  2\pi\times 30.2$~MHz. The red and green inset panels show the examples of two-mode clusters and three-mode clusters systems, where we highlight the spectral theoretical curves obtained from the model in Eq.~\eqref{Eq:ThreeModeH}, used to estimate the maximum \textit{intra-cluster} frequency separation around $\delta_{c}^{n,m} \approx 2\pi\times1.4$~MHz, which is much smaller than the \textit{inter-cluster} frequency gap $\Delta_{c,c+1}$. See~\cite{SM} for more details on the spectroscopy of the clusters systems considered in this work.}
	\label{Fig_Scheme}
\end{figure*}

\RefA{While collective behavior has been observed in optomechanical systems~\cite{Shkarin:14,Kharel:22,Barzanjeh:22} and superconducting circuit optomechanical platforms~\cite{Mahdi2004}, these experiments are largely performed in linear-response regimes or rely on dissipative preparation such as sideband cooling to the ground state. In contrast, collective effects at the single-excitation level in solid-state mechanical systems remain largely unexplored, particularly in the context of coherent quantum control and single-excitation dynamics.
In existing qubit–HBAR hybrid platforms~\cite{Chu:17,Chu:18,Crump:23,von_Lupke:24}, the large frequency spacing between mechanical modes and the relatively small qubit–mode coupling typically limit the interaction to a single near-resonant mode, thereby preventing the emergence of multimode collective behavior.
Going beyond the state of the art, we overcome these limitations by engineering the mode frequency spacing and enhancing the qubit–mode coupling strength, enabling strong and simultaneous coupling between a superconducting qubit and multiple near-resonant HBAR modes. As a result, collective effects naturally emerge at the single-excitation level, allowing coherent and efficient coupling of the qubit to multiple mechanical modes simultaneously. This regime, inaccessible in previous solid-state mechanical systems, provides a new route toward exploiting multimode collective dynamics for quantum information processing.

In this letter, we demonstrate strong coupling between a superconducting transmon qubit and a cluster of near-resonant mechanical modes, realizing a multimode interaction regime in an HBAR system. 
We engineered a device that exhibits a non-uniform distribution in frequency, leading to frequency-localized clusters of modes, while simultaneously enhancing the coupling strength between the qubit and the modes. We harness this unconventional HBAR system to observe the emergence of collective Dicke acoustodynamics, for a multimode system at single-excitation regime. The collectivity in our system is characterized in two complementary ways: by the $\sqrt{N}$-enhanced interaction strength and by the entanglement generation between the qubit and the multimode acoustic resonators. 
}Additionally, our system also allows us to observe the transition from a static collective dynamics to a dynamically induced timed-Dicke regime~\cite{Arno2022,Pennetta2022}. Although timed-Dicke states have been theoretically and experimentally investigated in atomic physics and quantum optics~\cite{Scully:09,Guerin:16,Rui:20,He:20,Liedl:23}, our work provides the first experimental evidence of such states in a solid-state hybrid quantum system. This is enabled by the clustered mode spectrum of our HBAR system, which naturally supports the formation of timed-Dicke–like collective excitations.


\textit{Clustered Modes in an Acoustic Mechanical Resonator.---} Our hybrid quantum acoustic device consists of a HBAR fabricated from a heterostructure composed of a 100-nm-thick molybdenum (Mo) electrode and a 900-nm-thick aluminum nitride (AlN) piezoelectric layer---see Figs.~\SubFig{Fig_Scheme}{a} and~\SubFig{Fig_Scheme}{b}. The structure is patterned into a circular disk with a radius of 55 $\mu$m. 
The top electrode of the HBAR is electrically connected to one of the capacitor pads of a superconducting transmon qubit. The qubit is a frequency-tunable transmon, enabling broadband spectroscopy across multiple acoustic modes. As illustrated in Fig.~\SubFig{Fig_Scheme}{b}, the qubit and HBAR are integrated using a flip-chip technique with a vertical separation of approximately 2 $\mu$m, which leads to an effective capacitive coupling strength $g$ ranging, approximately, from $ 2\pi\times0.53 $~MHz to $2\pi\times0.89$~MHz across different modes~\cite{SM}. 

As our first main result, the spectroscopic measurements of the qubit reveals that the HBAR acoustic spectrum exhibits the desired multimode cluster structure, as depicted in Fig.~\SubFig{Fig_Scheme}{c} and experimentally observed in Fig.~\SubFig{Fig_Scheme}{d}. Our HBAR system presents modes with frequency $\omega_{n,c}$, for the $n$-th mode of the $c$-th cluster, where each mode is coupled to the qubit with coupling strength $g_{n,c}$. The spectrum of the device is constituted by such clusters separated by \textit{inter-cluster} frequency detunings, $|\Delta_{c,c+1}|$, satisfying the dispersive regime, $|\Delta_{c,c+1}| \gg |g^{\mathrm{max}}_{c,c+1}|$, where $g^{\mathrm{max}}_{c,c+1} = \max_{n}[\{g_{n,c},g_{n,c+1}\}]$ denotes the strongest qubit-mode coupling within the neighboring clusters $c$ and $c+1$. However, these separated modes are not single modes like those ones found in HBARs~\cite{Chu:18,Crump:23,von_Lupke:24}, but rather clusters of closely spaced modes that can be treated as frequency-separated mode groups with a free spectral range (FSR) of approximately $2\pi\times30.2$~MHz. Each localized cluster consists of two or three modes separated by \textit{intra-cluster} detunings $\delta_{c}^{n,m  }$, roughly of order of $|\delta_{c}^{n,m}| \sim g_{n,c}$~\cite{SM}. 
Unlike the inter-cluster detuning $\Delta_{c,c+1}$, the intra-cluster detunings $\delta_{c}^{n,n+1}$ are on the order of the qubit-mode coupling strength. As a result, such modes in the clusters behave as near-degenerate modes.

\textit{Qubit-Clustered Modes Coupling.---} The theoretical model describing the system incorporates its key properties. First, the transmon, as an artifical atom, is capacitively coupled to the HBAR and operates in the strong coupling regime. 
Second, due to the intrinsic properties of HBAR systems, the modes within a given cluster $c$ constitute independent degrees of freedom and therefore do not interact with each other. As a result, the system can be described by an \textit{extended Jaynes-Cummings model} Hamiltonian of the form $\hat{H} = \hat{H}_\mathrm{0} + \hat{H}_{\mathrm{HBAR}} + \hat{H}_{\mathrm{int}}$, where counter-rotating terms are neglected, allowing the qubit-HBAR interaction to be expressed in the rotating wave approximation as 
\begin{equation}
	\hat{H}_{int}(t) = \sum\nolimits_{c=1}^{N}\sum\nolimits_{n=1}^{N_{c}}  \hbar g_{n,c} \left(\hat{f}_{\omega_{n},c}^{\dagger}\hat{a} + \hat{f}_{\omega_{n},c}\hat{a}^{\dagger}\right) , \label{Eq_Hamiltonian}
\end{equation}
where $H_{0} = \hbar\omega_\mathrm{qu}\hat{a}^{\dagger}\hat{a} + (\alpha /2)\hat{a}^{\dagger}\hat{a}^{\dagger}\hat{a}\hat{a}$ is the transmon qubit Hamiltonian, with creation, $\hat{a}^{\dagger}$,  annihilation, $\hat{a}$, operators and the qubit frequency $\omega_\mathrm{qu}$.  $\hat{H}_{\mathrm{HBAR}}=\sum_{c=1}^{N}\sum_{n=1}^{N_{c}}\omega_{n,c}\hbar \hat{f}_{\omega_{n},c}^{\dagger}\hat{f}_{\omega_{n},c}$ is the HBAR Hamiltonian, with $\hat{f}_{\omega_{n},c}$ and $\hat{f}^{\dagger}_{\omega_{n},c}$ are the annihilation and creation operators, respectively, of an phonon with frequency $\omega_{n,c}$ in the $c$-th phononic cluster constituted of $N_{c}$ modes.

\begin{figure}[t!]
	\includegraphics[width=\linewidth]{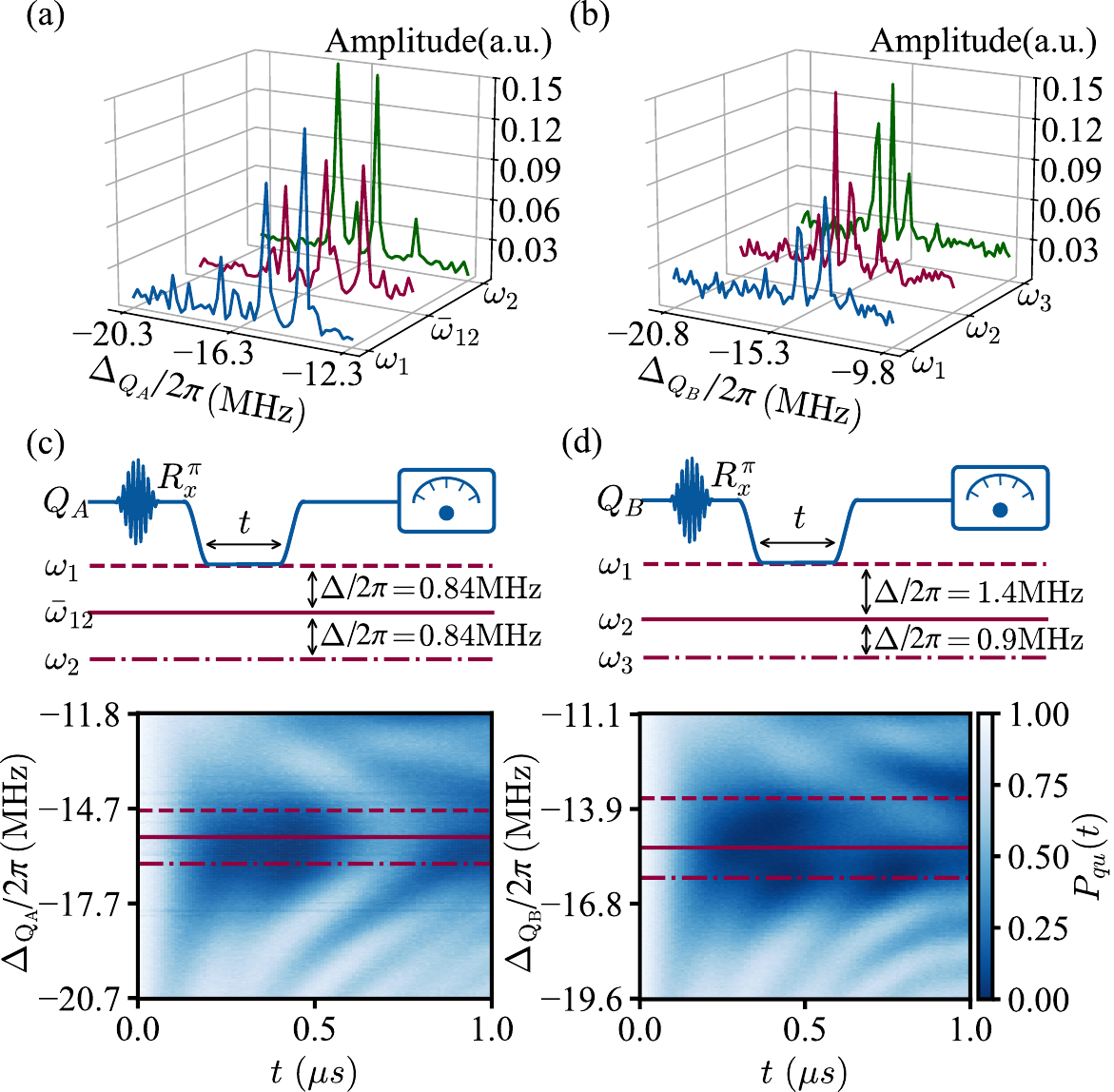}
	\caption{Coherence energy swapping between the qubit and multi-mode HBAR. (a) and (b) show qubit spectroscopy near the HBAR modes, where the qubit frequency  is tuned close to the HBAR modes, resulting in anti-crossings with three peaks (a) and four peaks (b), corresponding to the hybridization of the qubit with the two-mode HBAR in (a) and three modes HBAR in (b), respectively. (c) and (d) show the vacuum Rabi oscillations between the superconducting qubit and multimode HBAR. The experimental pulse sequence is shown in the top of (c) and (d), where the qubit is first excited and then tuned into resonance with the HBAR modes for a controlled interaction time $t$. The schematic illustrates coherent energy exchange between the qubit and two- or three-mode HBARs. The measured two-dimensional vacuum Rabi spectra in (c) and (d) reveal distinct features of the collective dynamics, consistent with the simulated circuit model shown in \cite{SM}. $Q_{A}$ and $Q_{B}$ refers to different qubits chosen from two devices of the hybrid quantum system, see Ref.~\cite{SM} further details.}
	\label{Fig_VacRabi}
\end{figure}

In Fig.~\SubFig{Fig_Scheme}{d} we show examples of the clusters in the spectroscopy. In the following we label the two-mode and three-mode clusters as $S_{2,c}$ and $S_{3,c}$ respectively. Due to the large inter-cluster frequency regime of our device, we can neglect the coupling with other modes in Eq.~\eqref{Eq_Hamiltonian}, and describe the dynamics with the approximate Hamiltonian for a cluster $S_{3,c}$ as
\begin{equation}
	\hat{H}_{c} = \hat{H}_\mathrm{qu} + \hat{H}_\mathrm{HBAR}^{c} + \hbar \sum\nolimits_{n=1}^{3}g_{n,c} \left(\hat{f}_{\omega_{n},c}^{\dagger}\hat{\sigma}^{-} + \hat{f}_{\omega_{n},c}\hat{\sigma}^{+}\right) , \label{Eq:ThreeModeH}
\end{equation}
where we define $\hat{H}_\mathrm{HBAR}^{c} = \hbar \sum_{n=1}^{3}\omega_{n,c} \hat{f}_{\omega_{n},c}^{\dagger}\hat{f}_{\omega_{n},c}$ and we assume the transmon as a two-level atom $\hat{H}_\mathrm{qu} = \hbar \omega_\mathrm{qu}\hat{\sigma}^{+}\hat{\sigma}^{-}$. This Hamiltonian is used to be confronted with experimental data, through exact diagonalisation of $\hat{H}_{c}$, as displayed in dash curves in the sub-figures of Fig.~\SubFig{Fig_Scheme}{d}. We compute the energy transitions between the ground state $\ket{\emptyset}\ket{g}$ and the single-excitation states of the complete Hamiltonian $\hat{H}_{c}$, taking into account the perturbed hybridized (dressed) states of the system. The agreement between the experiment and theoretical model reinforces the previous discussions. 

In Fig.~\ref{Fig_VacRabi}, these features are clearly reflected in both the spectroscopy and the time-domain measurements. As we tune the qubit across a mode cluster, the qubit spectroscopy exhibits three distinct peaks for the two-mode cluster and four peaks for the three-mode cluster as shown in Figs.~\SubFig{Fig_VacRabi}{a} and~\SubFig{Fig_VacRabi}{b}, respectively, consistent with the expected multimode hybridization. The time-domain measurements further capture the collective nature of the interaction. The vacuum Rabi oscillations in Figs.~\SubFig{Fig_VacRabi}{c} and ~\SubFig{Fig_VacRabi}{d} display pronounced interference patterns arising from the simultaneous coupling of the qubit to multiple closely spaced HBAR modes. These interference fringes provide direct signature of coherent energy exchange within the qubit–HBAR manifold and signify the emergence of collective dynamics arising from the hybridization of multiple modes.

\emph{Collective Acoustodynamics.} Different from other devices~\cite{Chu:18,Crump:23,von_Lupke:24}, the spectral structure of our system, and the non-dispersive regime of transmon-modes coupling, 
allow us to study for the first time the emergence of collective dynamics with acoustic modes. In fact, consider the transmon frequency around one of the modes $S_{3,c}$ and write the Hamiltonian $\hat{H}_{c}$ in the interaction picture as
\begin{equation}
	\hat{H}_{c}^{\mathrm{int}}(t) =  \hat{F}_{c}^{\dagger}(t)\hat{\sigma}^{-} + \hat{F}_{c}(t)\hat{\sigma}^{+}, 
\end{equation}
where $\hat{F}_{c}(t) = \sum_{n=1}^{3} g_{n,c} \hat{f}_{\omega_{n},c}e^{-i\Delta_{n,c}t}$ is the collective mode operators, and $\Delta_{n,c} = \omega_{n,c} - \omega_\mathrm{qu}$ is the detuning between the emitter and the $n$-th mode in cluster $c$. A similar Hamiltonian can be obtained for the two-mode case~\cite{SM}. From this equation, the collective aspect of the interaction becomes evident, as the atom transfers excitation to the phonon modes through the collective \textit{bright} mode $\hat{F}_{c}^{\dagger}(t)$. Moreover, it is immediate to observe the multiplexed aspect of the dynamics. In fact, by considering the system in the state $\ket{\emptyset}\ket{e}$ we get
\begin{equation}
	\hat{H}_{c}^{\mathrm{int}}(t)\ket{\emptyset}\ket{e} =  \left(\hat{F}_{c}^{\dagger}(t)\ket{\emptyset}\right)\left(\hat{\sigma}^{-}\ket{e} \right)
	= g_{\mathrm{eff},c}\ket{\Dcal_{c}(t)}\ket{g} , 
	\label{Eq:Hpsi}
\end{equation}
where we define the collective effective coupling $g_{\mathrm{eff},c} = \sqrt{G_{c}}$, with $G_{c} = \sum_{n}g_{n,c}^2$, and the dynamical \textit{timed-Dicke state}
\begin{equation}
	\ket{\Dcal_{c}(t)} =  \sum\nolimits_{n=1}^{N_{c}} \frac{g_{n,c}e^{i\Delta_{n,c}t}}{g_{\mathrm{eff},c}} \ket*{1_{\omega_{n},c}}.
    \label{Eq:Dt}
\end{equation}

From the above result, it is possible to observe that in the limit of $\Delta_{n,c}\rightarrow 0$, we obtain the \textit{static Dicke state} defined for our acoustic modes as $\ket*{\Dcal_{c,\mathrm{st}}} = \sum_{n} g_{n,c}\ket*{1_{\omega_{n},c}}/g_{\mathrm{eff},c} $. In particular, this also happens for the special case of degenerate modes, where all modes in a given cluster have the same frequency $\omega_{n,c}=\omega_{c}$. In this case, the local phase factors $e^{i\Delta_{n,c}t}$ become a \textit{global phase} $e^{i\Delta_{c} t}$, with $\Delta_{c} = \omega_{c} - \omega_\mathrm{qu}$. Therefore, when the qubit is resonant with the cluster modes ($\Delta_{c}=0$), the dynamics of the system is governed by the solution $\ket*{\psi_{\mathrm{D},c}(t)} = \cos(g_{\mathrm{eff},c} t)\ket{\emptyset,e} + \sin(g_{\mathrm{eff},c} t)\ket*{\Dcal_{c,\mathrm{st}},g}$.

\RefA{In conventional laser-cooled atomic ensembles~\cite{Scully:06,Yizun:20}, timed-Dicke states originate from spatial phase accumulation, $\vec{k}\!\cdot\!(\vec{r}_{j}-\vec{r}_{k})$ acquired as light with momentum $\vec{k}$ propagates between atoms located at positions $\vec{r}_{j}$ and $\vec{r}_{k}$. Accessing this regime typically requires inter-emitter separations comparable to the wavelength, which effectively suppresses collective interactions and approaches the single-emitter limit. In contrast, our platform enables the emergence of timed-Dicke dynamics through detuning-induced frequency dispersion of clustered modes $\omega_{n}$ over the course of the evolution of the system, while the system remains in the strong-coupling collective regime.} This mechanism, aligned with high controllability of the qubit, provides a versatile and controllable platform to explore the time continuous transition from static to timed-Dicke regime.

We consider frequency regimes containing two- to three-mode clusters and focus on short-time dynamical evolution to capture the transition from the static regime to the time-Dicke regime.
To this end, we compare the single-excitation transfer process in our system, which follows a timed-Dicke behavior, with the qubit population dynamics $P_{\mathrm{qu},N_{c}}^{\mathrm{col}}(t)$ of a theoretically predicted static Dicke regime, obtained from $\ket*{\psi_{\mathrm{D},c}(t)}$ \RefA{as 
$
P_{\mathrm{qu},N_{c}}^{\mathrm{col}}(t) = \cos^2(\sqrt{N_{c}}\bar{g}_{c} t)
$,
with $\bar{g}_{c} = \sum_{n=1}^{N_{c}} g_{n,c}/N_{c}$}~\footnote{\RefA{A more accurate estimate of the population dynamics $P_{\mathrm{qu},N_{c}}^{\mathrm{col}}(t)$ can be obtained} by considering the effective coupling $g_{\mathrm{eff},c}$ as defined in Eq.~\eqref{Eq:Hpsi}. However, the variation in individual coupling strengths prevents an explicit expression of the Dicke enhancement as $\sqrt{N_{c}}$. Using the average $\bar{g}_{c}$ provides a reliable estimate for the expected behavior.}.

\begin{figure}[t!]
	\includegraphics[width=\linewidth]{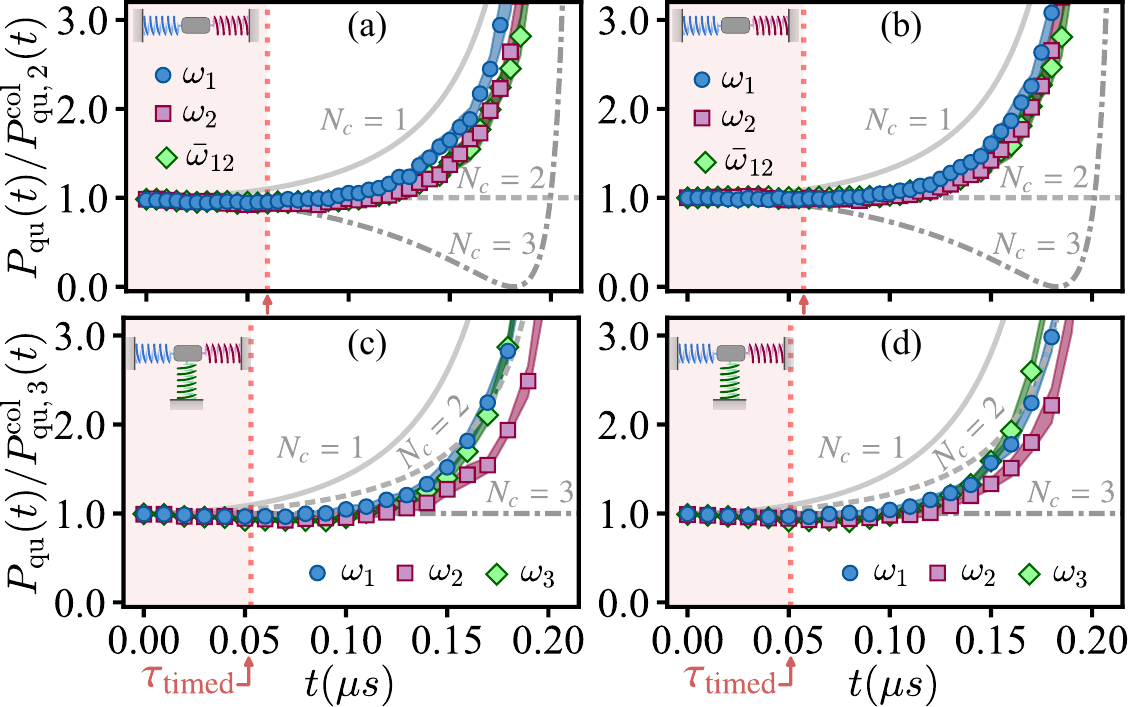}
	\caption{Experimentally measured qubit population, normalized by the collective population of the static Dicke dynamics $P_{\mathrm{qu},N_c}^{\mathrm{col}}(t)$, as function of the time to systems (a) $S_{2,1}$ and (b) $S_{2,2}$ with two modes, and the three mode systems (c) $S_{3,1}$ and (d) $S_{3,2}$. Gray curves describe $P_{\mathrm{qu},N_c}^{\mathrm{col}}(t)/P_{\mathrm{qu},2}^{\mathrm{col}}(t)$ in (a) and (b), and $P_{\mathrm{qu},N_c}^{\mathrm{col}}(t)/P_{\mathrm{qu},3}^{\mathrm{col}}(t)$ for (c) and (d). Error bars are shown as filled regions in the graphs.}
	\label{Fig:CollectiveBehavior}
\end{figure}

It is possible to distinguish the expected behaviors for different $N_c$, as shown in Fig.~\ref{Fig:CollectiveBehavior}, for the cases of two- and three-mode clusters available in our device. When contrasting the experimental data (symbols) with theoretical prediction (curves), the $\sqrt{N_{c}}$ enhancement in the population transfer rate is evident for the two- and three-mode case. However, significant deviation from the static Dicke collective behavior is observed after a time transition $\tau_{\mathrm{timed}}$. This divergence is due to the influence of the dynamical phases $\Delta_{n,c} t$ in Eq.~\eqref{Eq:Dt}, showing the transition between static to timed-Dicke regimes.

To obtain an estimate of the transition time for the $c$-th cluster, we compute the overlap between the static and timed-Dicke state as $\Fcal_{N_{c}}(t) = |\bra{\Dcal_{c,\mathrm{st}}}\ket{\Dcal_{c}(t)}|^2$ for a $N$-mode cluster, and we impose a minimum threshold fidelity $\mathcal{F}_{0}$ in which the system state remains approximately the same as the static Dicke state. Then, by assuming cluster modes with identical intra-cluster spacing, i.e. $\omega_{n+1,c}-\omega_{n,c} = \delta_{c}$, we can find the timed-Dicke transition time approximately as~\cite{SM}
\begin{equation}
	\tau_{\mathrm{timed}}(N_{c}) = \frac{1}{\delta_{c}}\frac{2 \sqrt{3-3 \mathcal{F}_{0}}}{\sqrt{N_{c}^2-1}} .
\end{equation}

This result demonstrates the complexity to observe static Dicke dynamics in these systems, as  $\tau_{\mathrm{timed}}$ becomes short for clusters with large number of non-degenerated modes. However, in our system, the average spectral frequency detuning between the modes allows us to observe a near-static-Dicke regime for a sufficient time $\tau_{\mathrm{timed}}$---highlighted in Fig.~\ref{Fig:CollectiveBehavior}.

\textit{Qubit and phononic-cluster entanglement.} According to first order expansion in Eq.~\eqref{Eq:Hpsi}, the qubit-modes collective excitation transfer creates entanglement between the qubit and the clustered modes. \RefA{To detect entanglement in the system}, first we restrict our evolution to short times where the system approximately evolves according to the Dicke evolution $\ket*{\psi_{\mathrm{D},c}(t)}$, \RefA{as it guarantees that the decoherence effects on the qubit are negligible and the system remains approximately pure in this timescale}. 

For pure states, or approximately pure, the entanglement between two parties, $A$ and $B$, can be directly \RefA{detected} by computing the purity $\Pcal (t) = \mathrm{tr}\big(\hat{\rho}_{A}^2(t)\big)$, where $\hat{\rho}_{A}(t)$ is the reduced density matrix of the part $A$. In our system, we have full access to the qubit information and therefore we consider $\hat{\rho}_{A}(t)=\hat{\rho}_{\mathrm{qu}}(t)$, with the qubit reduced density matrix $\hat{\rho}_{\mathrm{qu}}(t) = (1-|P_{\mathrm{qu}}^{\mathrm{col}}(N_{c})|^2)|\ket{g}\bra{g} + |P_{\mathrm{qu}}^{\mathrm{col}}(N_{c})|^2\ket{e}\bra{e}$ obtained from $\ket{\psi_{\mathrm{D},c}(t)}$~(see~\cite{SM} for details). Therefore, the purity reads
\begin{equation}
	\Pcal_{N_{c}} (t)= \frac{1}{4}\left[3 + \cos(4g_{\mathrm{eff},c}t)\right] \approx \frac{1}{4}\left[3 + \cos(4 \sqrt{N_{c}} \bar{g}_{c} t)\right] . \label{Eq:Purity}
\end{equation}

\begin{figure}[t!]
	\includegraphics[width=\linewidth]{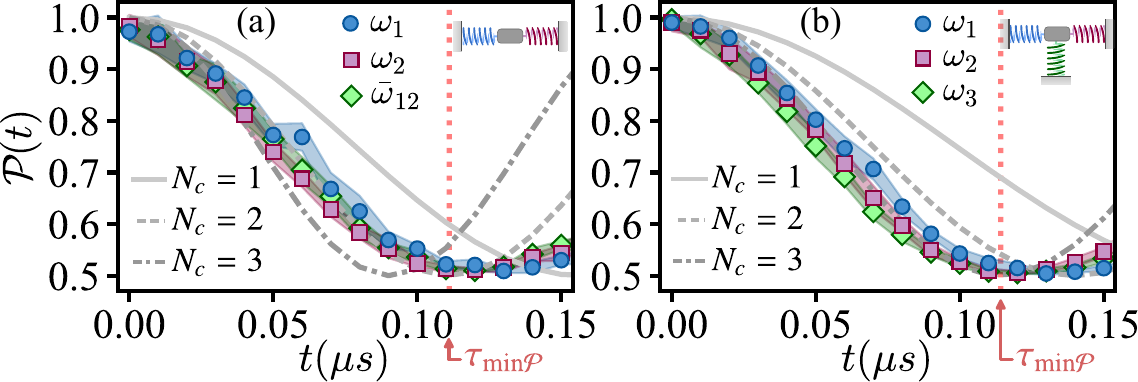}
	\caption{Qubit purity as function of time for (a) system $S_{2,1}$ and (b) system $S_{3,1}$, with curves showing the expected behavior of static Dicke regime for different $N_{c}$. The red dashed vertical line is the minimum purity time predicted by the analytical $\tau_{\mathrm{min}\Pcal}$. Error bars are shown as filled regions.}
	\label{Fig:Purity}
\end{figure}

In Figs.~\SubFig{Fig:Purity}{a} and ~\SubFig{Fig:Purity}{b} we present the purity dynamics for the cases in which the atom is coupled to two-mode and three-mode clusters, respectively. As suggested by Eq.~\eqref{Eq:Purity}, we are able to observe the effect of collectivity on the qubit-HBAR entanglement generation during the evolution, as the behavior of the purity approximately follows this solution. Furthermore, by computing the time interval $\tau_{\mathrm{min}\Pcal} = \pi /4 \sqrt{N_{c}} \bar{g}_{c}$ to reach the minimum of \RefA{the purity~\cite{SM}, we observe an} accurate agreement between the theoretical prediction---see horizontal dashed line in Fig.~\ref{Fig:Purity}---and the local minimum of the experimental data. These findings constitute a clear evidence of coherent and correlated qubit-HBAR excitation transfer, \RefA{indicating the emergence of entanglement that reveals the quantum nature of the collective motion.}

\emph{Conclusions.---} In this work we introduced fundamental principles of collective quantum vibrations of acoustic modes in a macroscopic mechanical resonator. By engineering of an unconventional hybrid quantum system---where a superconducting qubit is coupled to clusters with $N_{c}=2$ and $N_{c}=3$ acoustic modes of a macroscopic HBAR---we experimentally realized and characterized \RefA{collective Dicke dynamics} in the \RefA{single-excitation regime}. We showed that such a dynamics emerges mainly from the coupling of a qubit with near-resonant multiple mechanical modes whose frequency spacings are comparable to the qubit–mode coupling strength, leading to a collective dynamics in the mechanical cluster. In addition, time-domain measurements of the qubit’s purity \RefA{provides evidence for strong correlations between qubit and the multimode acoustic resonator consistent with entangled Dicke dynamics, revealing the quantum mechanical nature of the collective effects observed in our system.}

We further observe the transition from static to time-dependent Dicke regimes, consistent with both theoretical predictions and measured qubit population dynamics. The demonstrated crossover between \RefA{static and time-Dicke phases establishes this platform as a viable route to engineering non-equilibrium} collective dynamics. With modest scaling to entangle multimode clusters via the qubit, the same architecture could serve as a programmable analogue quantum simulator of many-body Dicke and structured spin-boson models.

The relevance of our results goes beyond the fundamental aspect of our work. For instance, the dynamical behavior of our system also open the door for potential applications in superconducting integrated devices. For instance, the timed-Dicke regime observed in our work could be used to the development of universal quantum erasers, as time-dependent dynamical phases develop an important role for efficient erasing of collective states in multiple-qubit systems~\cite{Diniz:25}. Potential applications in quantum encoding and information storage are also enabled by our system. For instance, Eq.~\eqref{Eq:Hpsi} demonstrates the ability to couple a qubit simultaneously to spectrally distinct mechanical modes is a key ingredient for microwave-phonon frequency-multiplexed transduction architectures and bosonic-entangled coding operations. Moreover, the superradiant nature of the qubit–mode population transfer into multiple HBAR modes can enable fast, superradiance-based, frequency-multiplexed phononic memory~\cite{Rastogi:22} for local and temporary information storage.

\begin{acknowledgments}
	\textit{Acknowledgments.--} DT acknowledges the support from Tobias Kippenberg, Jian-Qiang You and Yi Yin on this project. This work is supported by the National Natural Science Foundation of China (12574550, 11934010,12004167, 12205137), the Key-Area Research and Development Program of Guangdong Province (Grants No. 2018B030326001),  the Quantum Science and Technology-National Science and Technology Major Project (Grant No. 2021ZD0301703). ACS acknowledges support from the Comunidad de Madrid through the program Ayudas de Atracción de Talento Investigador ``César Nombela", under the grant No.~2024-T1/COM-31530 (Project SWiQL).
\end{acknowledgments}



\begin{thebibliography}{57}%
\makeatletter
\providecommand \@ifxundefined [1]{%
 \@ifx{#1\undefined}
}%
\providecommand \@ifnum [1]{%
 \ifnum #1\expandafter \@firstoftwo
 \else \expandafter \@secondoftwo
 \fi
}%
\providecommand \@ifx [1]{%
 \ifx #1\expandafter \@firstoftwo
 \else \expandafter \@secondoftwo
 \fi
}%
\providecommand \natexlab [1]{#1}%
\providecommand \enquote  [1]{``#1''}%
\providecommand \bibnamefont  [1]{#1}%
\providecommand \bibfnamefont [1]{#1}%
\providecommand \citenamefont [1]{#1}%
\providecommand \href@noop [0]{\@secondoftwo}%
\providecommand \href [0]{\begingroup \@sanitize@url \@href}%
\providecommand \@href[1]{\@@startlink{#1}\@@href}%
\providecommand \@@href[1]{\endgroup#1\@@endlink}%
\providecommand \@sanitize@url [0]{\catcode `\\12\catcode `\$12\catcode
  `\&12\catcode `\#12\catcode `\^12\catcode `\_12\catcode `\%12\relax}%
\providecommand \@@startlink[1]{}%
\providecommand \@@endlink[0]{}%
\providecommand \url  [0]{\begingroup\@sanitize@url \@url }%
\providecommand \@url [1]{\endgroup\@href {#1}{\urlprefix }}%
\providecommand \urlprefix  [0]{URL }%
\providecommand \Eprint [0]{\href }%
\providecommand \doibase [0]{https://doi.org/}%
\providecommand \selectlanguage [0]{\@gobble}%
\providecommand \bibinfo  [0]{\@secondoftwo}%
\providecommand \bibfield  [0]{\@secondoftwo}%
\providecommand \translation [1]{[#1]}%
\providecommand \BibitemOpen [0]{}%
\providecommand \bibitemStop [0]{}%
\providecommand \bibitemNoStop [0]{.\EOS\space}%
\providecommand \EOS [0]{\spacefactor3000\relax}%
\providecommand \BibitemShut  [1]{\csname bibitem#1\endcsname}%
\let\auto@bib@innerbib\@empty
\bibitem [{\citenamefont {Dicke}(1954)}]{Dick1954}%
  \BibitemOpen
  \bibfield  {author} {\bibinfo {author} {\bibfnamefont {R.~H.}\ \bibnamefont
  {Dicke}},\ }\bibfield  {title} {\bibinfo {title} {Coherence in spontaneous
  radiation processes},\ }\href {https://doi.org/10.1103/PhysRev.93.99}
  {\bibfield  {journal} {\bibinfo  {journal} {Phys. Rev.}\ }\textbf {\bibinfo
  {volume} {93}},\ \bibinfo {pages} {99} (\bibinfo {year} {1954})}\BibitemShut
  {NoStop}%
\bibitem [{Gro(1982)}]{Gross:82}%
  \BibitemOpen
  \bibfield  {title} {\bibinfo {title} {Superradiance: An essay on the theory
  of collective spontaneous emission},\ }\href@noop {} {\bibfield  {journal}
  {\bibinfo  {journal} {Physics Reports}\ }\textbf {\bibinfo {volume} {93}},\
  \bibinfo {pages} {301} (\bibinfo {year} {1982})}\BibitemShut {NoStop}%
\bibitem [{\citenamefont {Haake}\ \emph {et~al.}(1993)\citenamefont {Haake},
  \citenamefont {Kolobov}, \citenamefont {Fabre}, \citenamefont {Giacobino},\
  and\ \citenamefont {Reynaud}}]{Haake:93}%
  \BibitemOpen
  \bibfield  {author} {\bibinfo {author} {\bibfnamefont {F.}~\bibnamefont
  {Haake}}, \bibinfo {author} {\bibfnamefont {M.~I.}\ \bibnamefont {Kolobov}},
  \bibinfo {author} {\bibfnamefont {C.}~\bibnamefont {Fabre}}, \bibinfo
  {author} {\bibfnamefont {E.}~\bibnamefont {Giacobino}},\ and\ \bibinfo
  {author} {\bibfnamefont {S.}~\bibnamefont {Reynaud}},\ }\bibfield  {title}
  {\bibinfo {title} {Superradiant laser},\ }\href
  {https://doi.org/10.1103/PhysRevLett.71.995} {\bibfield  {journal} {\bibinfo
  {journal} {Phys. Rev. Lett.}\ }\textbf {\bibinfo {volume} {71}},\ \bibinfo
  {pages} {995} (\bibinfo {year} {1993})}\BibitemShut {NoStop}%
\bibitem [{\citenamefont {Bohnet}\ \emph {et~al.}(2012)\citenamefont {Bohnet},
  \citenamefont {Chen}, \citenamefont {Weiner}, \citenamefont {Meiser},
  \citenamefont {Holland},\ and\ \citenamefont {Thompson}}]{Bohnet:12}%
  \BibitemOpen
  \bibfield  {author} {\bibinfo {author} {\bibfnamefont {J.~G.}\ \bibnamefont
  {Bohnet}}, \bibinfo {author} {\bibfnamefont {Z.}~\bibnamefont {Chen}},
  \bibinfo {author} {\bibfnamefont {J.~M.}\ \bibnamefont {Weiner}}, \bibinfo
  {author} {\bibfnamefont {D.}~\bibnamefont {Meiser}}, \bibinfo {author}
  {\bibfnamefont {M.~J.}\ \bibnamefont {Holland}},\ and\ \bibinfo {author}
  {\bibfnamefont {J.~K.}\ \bibnamefont {Thompson}},\ }\bibfield  {title}
  {\bibinfo {title} {A steady-state superradiant laser with less than one
  intracavity photon},\ }\href
  {https://doi.org/https://doi.org/10.1038/nature10920} {\bibfield  {journal}
  {\bibinfo  {journal} {Nature}\ }\textbf {\bibinfo {volume} {484}},\ \bibinfo
  {pages} {78} (\bibinfo {year} {2012})}\BibitemShut {NoStop}%
\bibitem [{\citenamefont {Scully}(2015)}]{Scully:15}%
  \BibitemOpen
  \bibfield  {author} {\bibinfo {author} {\bibfnamefont {M.~O.}\ \bibnamefont
  {Scully}},\ }\bibfield  {title} {\bibinfo {title} {Single photon subradiance:
  Quantum control of spontaneous emission and ultrafast readout},\ }\href
  {https://doi.org/10.1103/PhysRevLett.115.243602} {\bibfield  {journal}
  {\bibinfo  {journal} {Phys. Rev. Lett.}\ }\textbf {\bibinfo {volume} {115}},\
  \bibinfo {pages} {243602} (\bibinfo {year} {2015})}\BibitemShut {NoStop}%
\bibitem [{\citenamefont {Perarnau-Llobet}\ \emph {et~al.}(2020)\citenamefont
  {Perarnau-Llobet}, \citenamefont {González-Tudela},\ and\ \citenamefont
  {Cirac}}]{Perarnau:20}%
  \BibitemOpen
  \bibfield  {author} {\bibinfo {author} {\bibfnamefont {M.}~\bibnamefont
  {Perarnau-Llobet}}, \bibinfo {author} {\bibfnamefont {A.}~\bibnamefont
  {González-Tudela}},\ and\ \bibinfo {author} {\bibfnamefont {J.~I.}\
  \bibnamefont {Cirac}},\ }\bibfield  {title} {\bibinfo {title} {Multimode fock
  states with large photon number: effective descriptions and applications in
  quantum metrology},\ }\href {https://doi.org/10.1088/2058-9565/ab6ce5}
  {\bibfield  {journal} {\bibinfo  {journal} {Quantum Science and Technology}\
  }\textbf {\bibinfo {volume} {5}},\ \bibinfo {pages} {025003} (\bibinfo {year}
  {2020})}\BibitemShut {NoStop}%
\bibitem [{\citenamefont {Sheremet}\ \emph {et~al.}(2023)\citenamefont
  {Sheremet}, \citenamefont {Petrov}, \citenamefont {Iorsh}, \citenamefont
  {Poshakinskiy},\ and\ \citenamefont {Poddubny}}]{Sheremet:23}%
  \BibitemOpen
  \bibfield  {author} {\bibinfo {author} {\bibfnamefont {A.~S.}\ \bibnamefont
  {Sheremet}}, \bibinfo {author} {\bibfnamefont {M.~I.}\ \bibnamefont
  {Petrov}}, \bibinfo {author} {\bibfnamefont {I.~V.}\ \bibnamefont {Iorsh}},
  \bibinfo {author} {\bibfnamefont {A.~V.}\ \bibnamefont {Poshakinskiy}},\ and\
  \bibinfo {author} {\bibfnamefont {A.~N.}\ \bibnamefont {Poddubny}},\
  }\bibfield  {title} {\bibinfo {title} {Waveguide quantum electrodynamics:
  Collective radiance and photon-photon correlations},\ }\href
  {https://doi.org/10.1103/RevModPhys.95.015002} {\bibfield  {journal}
  {\bibinfo  {journal} {Rev. Mod. Phys.}\ }\textbf {\bibinfo {volume} {95}},\
  \bibinfo {pages} {015002} (\bibinfo {year} {2023})}\BibitemShut {NoStop}%
\bibitem [{\citenamefont {Ritsch}\ \emph {et~al.}(2013)\citenamefont {Ritsch},
  \citenamefont {Domokos}, \citenamefont {Brennecke},\ and\ \citenamefont
  {Esslinger}}]{Ritsch2013}%
  \BibitemOpen
  \bibfield  {author} {\bibinfo {author} {\bibfnamefont {H.}~\bibnamefont
  {Ritsch}}, \bibinfo {author} {\bibfnamefont {P.}~\bibnamefont {Domokos}},
  \bibinfo {author} {\bibfnamefont {F.}~\bibnamefont {Brennecke}},\ and\
  \bibinfo {author} {\bibfnamefont {T.}~\bibnamefont {Esslinger}},\ }\bibfield
  {title} {\bibinfo {title} {Cold atoms in cavity-generated dynamical optical
  potentials},\ }\href {https://doi.org/10.1103/RevModPhys.85.553} {\bibfield
  {journal} {\bibinfo  {journal} {Rev. Mod. Phys.}\ }\textbf {\bibinfo {volume}
  {85}},\ \bibinfo {pages} {553} (\bibinfo {year} {2013})}\BibitemShut
  {NoStop}%
\bibitem [{\citenamefont {Chegnizadeh}\ \emph {et~al.}(2024)\citenamefont
  {Chegnizadeh}, \citenamefont {Scigliuzzo}, \citenamefont {Youssefi},
  \citenamefont {Kono}, \citenamefont {Guzovskii},\ and\ \citenamefont
  {Kippenberg}}]{Mahdi2004}%
  \BibitemOpen
  \bibfield  {author} {\bibinfo {author} {\bibfnamefont {M.}~\bibnamefont
  {Chegnizadeh}}, \bibinfo {author} {\bibfnamefont {M.}~\bibnamefont
  {Scigliuzzo}}, \bibinfo {author} {\bibfnamefont {A.}~\bibnamefont
  {Youssefi}}, \bibinfo {author} {\bibfnamefont {S.}~\bibnamefont {Kono}},
  \bibinfo {author} {\bibfnamefont {E.}~\bibnamefont {Guzovskii}},\ and\
  \bibinfo {author} {\bibfnamefont {T.~J.}\ \bibnamefont {Kippenberg}},\
  }\bibfield  {title} {\bibinfo {title} {Quantum collective motion of
  macroscopic mechanical oscillators},\ }\href
  {https://doi.org/10.1126/science.adr8187} {\bibfield  {journal} {\bibinfo
  {journal} {Science}\ }\textbf {\bibinfo {volume} {386}},\ \bibinfo {pages}
  {1383} (\bibinfo {year} {2024})}\BibitemShut {NoStop}%
\bibitem [{\citenamefont {Schmolke}\ and\ \citenamefont
  {Lutz}(2022)}]{Finn2022}%
  \BibitemOpen
  \bibfield  {author} {\bibinfo {author} {\bibfnamefont {F.}~\bibnamefont
  {Schmolke}}\ and\ \bibinfo {author} {\bibfnamefont {E.}~\bibnamefont
  {Lutz}},\ }\bibfield  {title} {\bibinfo {title} {Noise-induced quantum
  synchronization},\ }\href {https://doi.org/10.1103/PhysRevLett.129.250601}
  {\bibfield  {journal} {\bibinfo  {journal} {Phys. Rev. Lett.}\ }\textbf
  {\bibinfo {volume} {129}},\ \bibinfo {pages} {250601} (\bibinfo {year}
  {2022})}\BibitemShut {NoStop}%
\bibitem [{\citenamefont {Tao}\ \emph {et~al.}(2025)\citenamefont {Tao},
  \citenamefont {Schmolke}, \citenamefont {Hu}, \citenamefont {Huang},
  \citenamefont {Zhou}, \citenamefont {Zhang}, \citenamefont {Chu},
  \citenamefont {Zhang}, \citenamefont {Sun}, \citenamefont {Guo},
  \citenamefont {Niu}, \citenamefont {Weng}, \citenamefont {Liu}, \citenamefont
  {Zhong}, \citenamefont {Tan}, \citenamefont {Yu},\ and\ \citenamefont
  {Lutz}}]{Tao2025}%
  \BibitemOpen
  \bibfield  {author} {\bibinfo {author} {\bibfnamefont {Z.}~\bibnamefont
  {Tao}}, \bibinfo {author} {\bibfnamefont {F.}~\bibnamefont {Schmolke}},
  \bibinfo {author} {\bibfnamefont {C.-K.}\ \bibnamefont {Hu}}, \bibinfo
  {author} {\bibfnamefont {W.}~\bibnamefont {Huang}}, \bibinfo {author}
  {\bibfnamefont {Y.}~\bibnamefont {Zhou}}, \bibinfo {author} {\bibfnamefont
  {J.}~\bibnamefont {Zhang}}, \bibinfo {author} {\bibfnamefont
  {J.}~\bibnamefont {Chu}}, \bibinfo {author} {\bibfnamefont {L.}~\bibnamefont
  {Zhang}}, \bibinfo {author} {\bibfnamefont {X.}~\bibnamefont {Sun}}, \bibinfo
  {author} {\bibfnamefont {Z.}~\bibnamefont {Guo}}, \bibinfo {author}
  {\bibfnamefont {J.}~\bibnamefont {Niu}}, \bibinfo {author} {\bibfnamefont
  {W.}~\bibnamefont {Weng}}, \bibinfo {author} {\bibfnamefont {S.}~\bibnamefont
  {Liu}}, \bibinfo {author} {\bibfnamefont {Y.}~\bibnamefont {Zhong}}, \bibinfo
  {author} {\bibfnamefont {D.}~\bibnamefont {Tan}}, \bibinfo {author}
  {\bibfnamefont {D.}~\bibnamefont {Yu}},\ and\ \bibinfo {author}
  {\bibfnamefont {E.}~\bibnamefont {Lutz}},\ }\bibfield  {title} {\bibinfo
  {title} {Noise-induced quantum synchronization with entangled oscillations},\
  }\href {https://doi.org/doi.org/10.1038/s41467-025-63196-6} {\bibfield
  {journal} {\bibinfo  {journal} {Nature Communications}\ }\textbf {\bibinfo
  {volume} {16}},\ \bibinfo {pages} {8457} (\bibinfo {year}
  {2025})}\BibitemShut {NoStop}%
\bibitem [{\citenamefont {Anderson}\ \emph {et~al.}(1995)\citenamefont
  {Anderson}, \citenamefont {Ensher}, \citenamefont {Matthews}, \citenamefont
  {Wieman},\ and\ \citenamefont {Cornell}}]{Anderson1995}%
  \BibitemOpen
  \bibfield  {author} {\bibinfo {author} {\bibfnamefont {M.~H.}\ \bibnamefont
  {Anderson}}, \bibinfo {author} {\bibfnamefont {J.~R.}\ \bibnamefont
  {Ensher}}, \bibinfo {author} {\bibfnamefont {M.~R.}\ \bibnamefont
  {Matthews}}, \bibinfo {author} {\bibfnamefont {C.~E.}\ \bibnamefont
  {Wieman}},\ and\ \bibinfo {author} {\bibfnamefont {E.~A.}\ \bibnamefont
  {Cornell}},\ }\bibfield  {title} {\bibinfo {title} {Observation of
  bose-einstein condensation in a dilute atomic vapor},\ }\href@noop {}
  {\bibfield  {journal} {\bibinfo  {journal} {Science}\ }\textbf {\bibinfo
  {volume} {269}},\ \bibinfo {pages} {198} (\bibinfo {year}
  {1995})}\BibitemShut {NoStop}%
\bibitem [{\citenamefont {Fink}\ \emph {et~al.}(2009)\citenamefont {Fink},
  \citenamefont {Bianchetti}, \citenamefont {Baur}, \citenamefont {G\"oppl},
  \citenamefont {Steffen}, \citenamefont {Filipp}, \citenamefont {Leek},
  \citenamefont {Blais},\ and\ \citenamefont {Wallraff}}]{Fink2009}%
  \BibitemOpen
  \bibfield  {author} {\bibinfo {author} {\bibfnamefont {J.~M.}\ \bibnamefont
  {Fink}}, \bibinfo {author} {\bibfnamefont {R.}~\bibnamefont {Bianchetti}},
  \bibinfo {author} {\bibfnamefont {M.}~\bibnamefont {Baur}}, \bibinfo {author}
  {\bibfnamefont {M.}~\bibnamefont {G\"oppl}}, \bibinfo {author} {\bibfnamefont
  {L.}~\bibnamefont {Steffen}}, \bibinfo {author} {\bibfnamefont
  {S.}~\bibnamefont {Filipp}}, \bibinfo {author} {\bibfnamefont {P.~J.}\
  \bibnamefont {Leek}}, \bibinfo {author} {\bibfnamefont {A.}~\bibnamefont
  {Blais}},\ and\ \bibinfo {author} {\bibfnamefont {A.}~\bibnamefont
  {Wallraff}},\ }\bibfield  {title} {\bibinfo {title} {Dressed collective qubit
  states and the tavis-cummings model in circuit qed},\ }\href
  {https://doi.org/10.1103/PhysRevLett.103.083601} {\bibfield  {journal}
  {\bibinfo  {journal} {Phys. Rev. Lett.}\ }\textbf {\bibinfo {volume} {103}},\
  \bibinfo {pages} {083601} (\bibinfo {year} {2009})}\BibitemShut {NoStop}%
\bibitem [{\citenamefont {Wang}\ \emph {et~al.}(2020)\citenamefont {Wang},
  \citenamefont {Li}, \citenamefont {Feng}, \citenamefont {Song}, \citenamefont
  {Song}, \citenamefont {Liu}, \citenamefont {Guo}, \citenamefont {Zhang},
  \citenamefont {Dong}, \citenamefont {Zheng}, \citenamefont {Wang},\ and\
  \citenamefont {Wang}}]{Wang:20}%
  \BibitemOpen
  \bibfield  {author} {\bibinfo {author} {\bibfnamefont {Z.}~\bibnamefont
  {Wang}}, \bibinfo {author} {\bibfnamefont {H.}~\bibnamefont {Li}}, \bibinfo
  {author} {\bibfnamefont {W.}~\bibnamefont {Feng}}, \bibinfo {author}
  {\bibfnamefont {X.}~\bibnamefont {Song}}, \bibinfo {author} {\bibfnamefont
  {C.}~\bibnamefont {Song}}, \bibinfo {author} {\bibfnamefont {W.}~\bibnamefont
  {Liu}}, \bibinfo {author} {\bibfnamefont {Q.}~\bibnamefont {Guo}}, \bibinfo
  {author} {\bibfnamefont {X.}~\bibnamefont {Zhang}}, \bibinfo {author}
  {\bibfnamefont {H.}~\bibnamefont {Dong}}, \bibinfo {author} {\bibfnamefont
  {D.}~\bibnamefont {Zheng}}, \bibinfo {author} {\bibfnamefont
  {H.}~\bibnamefont {Wang}},\ and\ \bibinfo {author} {\bibfnamefont {D.-W.}\
  \bibnamefont {Wang}},\ }\bibfield  {title} {\bibinfo {title} {Controllable
  switching between superradiant and subradiant states in a 10-qubit
  superconducting circuit},\ }\href
  {https://doi.org/10.1103/PhysRevLett.124.013601} {\bibfield  {journal}
  {\bibinfo  {journal} {Phys. Rev. Lett.}\ }\textbf {\bibinfo {volume} {124}},\
  \bibinfo {pages} {013601} (\bibinfo {year} {2020})}\BibitemShut {NoStop}%
\bibitem [{\citenamefont {Clerk}\ \emph {et~al.}(2020)\citenamefont {Clerk},
  \citenamefont {Lehnert}, \citenamefont {Bertet}, \citenamefont {Petta},\ and\
  \citenamefont {Nakamura}}]{Hybrid20}%
  \BibitemOpen
  \bibfield  {author} {\bibinfo {author} {\bibfnamefont {A.~A.}\ \bibnamefont
  {Clerk}}, \bibinfo {author} {\bibfnamefont {K.~W.}\ \bibnamefont {Lehnert}},
  \bibinfo {author} {\bibfnamefont {P.}~\bibnamefont {Bertet}}, \bibinfo
  {author} {\bibfnamefont {J.~R.}\ \bibnamefont {Petta}},\ and\ \bibinfo
  {author} {\bibfnamefont {Y.}~\bibnamefont {Nakamura}},\ }\bibfield  {title}
  {\bibinfo {title} {Hybrid quantum systems with circuit quantum
  electrodynamics},\ }\href
  {https://doi.org/https://doi.org/10.1038/s41567-020-0797-9} {\bibfield
  {journal} {\bibinfo  {journal} {Nature Physics}\ }\textbf {\bibinfo {volume}
  {16}},\ \bibinfo {pages} {1745} (\bibinfo {year} {2020})}\BibitemShut
  {NoStop}%
\bibitem [{\citenamefont {Chu}\ and\ \citenamefont
  {Gröblacher}(2020)}]{Chu2020APL}%
  \BibitemOpen
  \bibfield  {author} {\bibinfo {author} {\bibfnamefont {Y.}~\bibnamefont
  {Chu}}\ and\ \bibinfo {author} {\bibfnamefont {S.}~\bibnamefont
  {Gröblacher}},\ }\bibfield  {title} {\bibinfo {title} {A perspective on
  hybrid quantum opto- and electromechanical systems},\ }\href
  {https://doi.org/10.1063/5.0021088} {\bibfield  {journal} {\bibinfo
  {journal} {Applied Physics Letters}\ }\textbf {\bibinfo {volume} {117}},\
  \bibinfo {pages} {150503} (\bibinfo {year} {2020})}\BibitemShut {NoStop}%
\bibitem [{\citenamefont {O’Connell}\ \emph {et~al.}(2010)\citenamefont
  {O’Connell}, \citenamefont {Hofheinz}, \citenamefont {Ansmann},
  \citenamefont {Bialczak}, \citenamefont {Lenander}, \citenamefont {Lucero},
  \citenamefont {Neeley}, \citenamefont {Sank}, \citenamefont {Wang},
  \citenamefont {Weides}, \citenamefont {Wenner}, \citenamefont {Martinis},\
  and\ \citenamefont {Cleland}}]{Connell10}%
  \BibitemOpen
  \bibfield  {author} {\bibinfo {author} {\bibfnamefont {A.~D.}\ \bibnamefont
  {O’Connell}}, \bibinfo {author} {\bibfnamefont {M.}~\bibnamefont
  {Hofheinz}}, \bibinfo {author} {\bibfnamefont {M.}~\bibnamefont {Ansmann}},
  \bibinfo {author} {\bibfnamefont {R.~C.}\ \bibnamefont {Bialczak}}, \bibinfo
  {author} {\bibfnamefont {M.}~\bibnamefont {Lenander}}, \bibinfo {author}
  {\bibfnamefont {E.}~\bibnamefont {Lucero}}, \bibinfo {author} {\bibfnamefont
  {M.}~\bibnamefont {Neeley}}, \bibinfo {author} {\bibfnamefont
  {D.}~\bibnamefont {Sank}}, \bibinfo {author} {\bibfnamefont {H.}~\bibnamefont
  {Wang}}, \bibinfo {author} {\bibfnamefont {M.}~\bibnamefont {Weides}},
  \bibinfo {author} {\bibfnamefont {J.}~\bibnamefont {Wenner}}, \bibinfo
  {author} {\bibfnamefont {J.~M.}\ \bibnamefont {Martinis}},\ and\ \bibinfo
  {author} {\bibfnamefont {A.~N.}\ \bibnamefont {Cleland}},\ }\bibfield
  {title} {\bibinfo {title} {Quantum ground state and single-phonon control of
  a mechanical resonator},\ }\href
  {https://doi.org/https://doi.org/10.1038/nature08967} {\bibfield  {journal}
  {\bibinfo  {journal} {Nature}\ }\textbf {\bibinfo {volume} {464}},\ \bibinfo
  {pages} {697} (\bibinfo {year} {2010})}\BibitemShut {NoStop}%
\bibitem [{\citenamefont {Teufel}\ \emph {et~al.}(2011)\citenamefont {Teufel},
  \citenamefont {Donner}, \citenamefont {Li}, \citenamefont {Harlow},
  \citenamefont {Allman}, \citenamefont {Cicak}, \citenamefont {Sirois},
  \citenamefont {Whittaker}, \citenamefont {Lehnert},\ and\ \citenamefont
  {Simmonds}}]{Teufel11}%
  \BibitemOpen
  \bibfield  {author} {\bibinfo {author} {\bibfnamefont {J.~D.}\ \bibnamefont
  {Teufel}}, \bibinfo {author} {\bibfnamefont {T.}~\bibnamefont {Donner}},
  \bibinfo {author} {\bibfnamefont {D.}~\bibnamefont {Li}}, \bibinfo {author}
  {\bibfnamefont {J.~W.}\ \bibnamefont {Harlow}}, \bibinfo {author}
  {\bibfnamefont {M.~S.}\ \bibnamefont {Allman}}, \bibinfo {author}
  {\bibfnamefont {K.}~\bibnamefont {Cicak}}, \bibinfo {author} {\bibfnamefont
  {A.~J.}\ \bibnamefont {Sirois}}, \bibinfo {author} {\bibfnamefont {J.~D.}\
  \bibnamefont {Whittaker}}, \bibinfo {author} {\bibfnamefont {K.~W.}\
  \bibnamefont {Lehnert}},\ and\ \bibinfo {author} {\bibfnamefont {R.~W.}\
  \bibnamefont {Simmonds}},\ }\bibfield  {title} {\bibinfo {title} {Sideband
  cooling of micromechanical motion to the quantum ground state},\ }\href
  {https://doi.org/https://doi.org/10.1038/nature10261} {\bibfield  {journal}
  {\bibinfo  {journal} {Nature}\ }\textbf {\bibinfo {volume} {475}},\ \bibinfo
  {pages} {359} (\bibinfo {year} {2011})}\BibitemShut {NoStop}%
\bibitem [{\citenamefont {Gustafsson}\ \emph {et~al.}(2014)\citenamefont
  {Gustafsson}, \citenamefont {Aref}, \citenamefont {Kockum}, \citenamefont
  {Ekström}, \citenamefont {Johansson},\ and\ \citenamefont
  {Delsing}}]{Martin14}%
  \BibitemOpen
  \bibfield  {author} {\bibinfo {author} {\bibfnamefont {M.~V.}\ \bibnamefont
  {Gustafsson}}, \bibinfo {author} {\bibfnamefont {T.}~\bibnamefont {Aref}},
  \bibinfo {author} {\bibfnamefont {A.~F.}\ \bibnamefont {Kockum}}, \bibinfo
  {author} {\bibfnamefont {M.~K.}\ \bibnamefont {Ekström}}, \bibinfo {author}
  {\bibfnamefont {G.}~\bibnamefont {Johansson}},\ and\ \bibinfo {author}
  {\bibfnamefont {P.}~\bibnamefont {Delsing}},\ }\bibfield  {title} {\bibinfo
  {title} {Propagating phonons coupled to an artificial atom},\ }\href
  {https://doi.org/10.1126/science.1257219} {\bibfield  {journal} {\bibinfo
  {journal} {Science}\ }\textbf {\bibinfo {volume} {346}},\ \bibinfo {pages}
  {207} (\bibinfo {year} {2014})}\BibitemShut {NoStop}%
\bibitem [{\citenamefont {Manenti}\ \emph {et~al.}(2017)\citenamefont
  {Manenti}, \citenamefont {Kockum}, \citenamefont {Patterson}, \citenamefont
  {Behrle}, \citenamefont {Rahamim}, \citenamefont {Tancredi}, \citenamefont
  {Nori},\ and\ \citenamefont {Leek}}]{Manenti:17}%
  \BibitemOpen
  \bibfield  {author} {\bibinfo {author} {\bibfnamefont {R.}~\bibnamefont
  {Manenti}}, \bibinfo {author} {\bibfnamefont {A.~F.}\ \bibnamefont {Kockum}},
  \bibinfo {author} {\bibfnamefont {A.}~\bibnamefont {Patterson}}, \bibinfo
  {author} {\bibfnamefont {T.}~\bibnamefont {Behrle}}, \bibinfo {author}
  {\bibfnamefont {J.}~\bibnamefont {Rahamim}}, \bibinfo {author} {\bibfnamefont
  {G.}~\bibnamefont {Tancredi}}, \bibinfo {author} {\bibfnamefont
  {F.}~\bibnamefont {Nori}},\ and\ \bibinfo {author} {\bibfnamefont {P.~J.}\
  \bibnamefont {Leek}},\ }\bibfield  {title} {\bibinfo {title} {Circuit quantum
  acoustodynamics with surface acoustic waves},\ }\href
  {https://doi.org/https://doi.org/10.1038/s41467-017-01063-9} {\bibfield
  {journal} {\bibinfo  {journal} {Nature communications}\ }\textbf {\bibinfo
  {volume} {8}},\ \bibinfo {pages} {975} (\bibinfo {year} {2017})}\BibitemShut
  {NoStop}%
\bibitem [{\citenamefont {Chu}\ \emph {et~al.}(2017)\citenamefont {Chu},
  \citenamefont {Kharel}, \citenamefont {Renninger}, \citenamefont {Burkhart},
  \citenamefont {Frunzio}, \citenamefont {Rakich},\ and\ \citenamefont
  {Schoelkopf}}]{Chu:17}%
  \BibitemOpen
  \bibfield  {author} {\bibinfo {author} {\bibfnamefont {Y.}~\bibnamefont
  {Chu}}, \bibinfo {author} {\bibfnamefont {P.}~\bibnamefont {Kharel}},
  \bibinfo {author} {\bibfnamefont {W.~H.}\ \bibnamefont {Renninger}}, \bibinfo
  {author} {\bibfnamefont {L.~D.}\ \bibnamefont {Burkhart}}, \bibinfo {author}
  {\bibfnamefont {L.}~\bibnamefont {Frunzio}}, \bibinfo {author} {\bibfnamefont
  {P.~T.}\ \bibnamefont {Rakich}},\ and\ \bibinfo {author} {\bibfnamefont
  {R.~J.}\ \bibnamefont {Schoelkopf}},\ }\bibfield  {title} {\bibinfo {title}
  {Quantum acoustics with superconducting qubits},\ }\href
  {https://doi.org/10.1126/science.aao1511} {\bibfield  {journal} {\bibinfo
  {journal} {Science}\ }\textbf {\bibinfo {volume} {358}},\ \bibinfo {pages}
  {199} (\bibinfo {year} {2017})}\BibitemShut {NoStop}%
\bibitem [{\citenamefont {Satzinger}\ \emph {et~al.}(2018)\citenamefont
  {Satzinger}, \citenamefont {Zhong}, \citenamefont {Chang}, \citenamefont
  {Peairs}, \citenamefont {Bienfait}, \citenamefont {Chou}, \citenamefont
  {Cleland}, \citenamefont {Conner}, \citenamefont {Dumur}, \citenamefont
  {Grebel} \emph {et~al.}}]{Satzinger18}%
  \BibitemOpen
  \bibfield  {author} {\bibinfo {author} {\bibfnamefont {K.~J.}\ \bibnamefont
  {Satzinger}}, \bibinfo {author} {\bibfnamefont {Y.}~\bibnamefont {Zhong}},
  \bibinfo {author} {\bibfnamefont {H.-S.}\ \bibnamefont {Chang}}, \bibinfo
  {author} {\bibfnamefont {G.~A.}\ \bibnamefont {Peairs}}, \bibinfo {author}
  {\bibfnamefont {A.}~\bibnamefont {Bienfait}}, \bibinfo {author}
  {\bibfnamefont {M.-H.}\ \bibnamefont {Chou}}, \bibinfo {author}
  {\bibfnamefont {A.}~\bibnamefont {Cleland}}, \bibinfo {author} {\bibfnamefont
  {C.~R.}\ \bibnamefont {Conner}}, \bibinfo {author} {\bibfnamefont
  {{\'E}.}~\bibnamefont {Dumur}}, \bibinfo {author} {\bibfnamefont
  {J.}~\bibnamefont {Grebel}}, \emph {et~al.},\ }\bibfield  {title} {\bibinfo
  {title} {Quantum control of surface acoustic-wave phonons},\ }\href
  {https://doi.org/10.1038/s41586-018-0719-5} {\bibfield  {journal} {\bibinfo
  {journal} {Nature}\ }\textbf {\bibinfo {volume} {563}},\ \bibinfo {pages}
  {661} (\bibinfo {year} {2018})}\BibitemShut {NoStop}%
\bibitem [{\citenamefont {Andersson}\ \emph {et~al.}(2019)\citenamefont
  {Andersson}, \citenamefont {Suri}, \citenamefont {Guo}, \citenamefont
  {Aref},\ and\ \citenamefont {Delsing}}]{Andersson19}%
  \BibitemOpen
  \bibfield  {author} {\bibinfo {author} {\bibfnamefont {G.}~\bibnamefont
  {Andersson}}, \bibinfo {author} {\bibfnamefont {B.}~\bibnamefont {Suri}},
  \bibinfo {author} {\bibfnamefont {L.}~\bibnamefont {Guo}}, \bibinfo {author}
  {\bibfnamefont {T.}~\bibnamefont {Aref}},\ and\ \bibinfo {author}
  {\bibfnamefont {P.}~\bibnamefont {Delsing}},\ }\bibfield  {title} {\bibinfo
  {title} {Non-exponential decay of a giant artificial atom},\ }\href@noop {}
  {\bibfield  {journal} {\bibinfo  {journal} {Nature Physics}\ }\textbf
  {\bibinfo {volume} {15}},\ \bibinfo {pages} {1123} (\bibinfo {year}
  {2019})}\BibitemShut {NoStop}%
\bibitem [{\citenamefont {Wollack}\ \emph {et~al.}(2022)\citenamefont
  {Wollack}, \citenamefont {Cleland}, \citenamefont {Gruenke}, \citenamefont
  {Wang}, \citenamefont {Arrangoiz-Arriola},\ and\ \citenamefont
  {Safavi-Naeini}}]{Wollack2022}%
  \BibitemOpen
  \bibfield  {author} {\bibinfo {author} {\bibfnamefont {E.~A.}\ \bibnamefont
  {Wollack}}, \bibinfo {author} {\bibfnamefont {A.~Y.}\ \bibnamefont
  {Cleland}}, \bibinfo {author} {\bibfnamefont {R.~G.}\ \bibnamefont
  {Gruenke}}, \bibinfo {author} {\bibfnamefont {Z.}~\bibnamefont {Wang}},
  \bibinfo {author} {\bibfnamefont {P.}~\bibnamefont {Arrangoiz-Arriola}},\
  and\ \bibinfo {author} {\bibfnamefont {A.~H.}\ \bibnamefont
  {Safavi-Naeini}},\ }\bibfield  {title} {\bibinfo {title} {Quantum state
  preparation and tomography of entangled mechanical resonators},\ }\href
  {https://doi.org/10.1038/s41586-022-04500-y} {\bibfield  {journal} {\bibinfo
  {journal} {Nature}\ }\textbf {\bibinfo {volume} {604}},\ \bibinfo {pages}
  {463} (\bibinfo {year} {2022})}\BibitemShut {NoStop}%
\bibitem [{\citenamefont {Arrangoiz-Arriola}\ \emph {et~al.}(2019)\citenamefont
  {Arrangoiz-Arriola}, \citenamefont {Wollack}, \citenamefont {Wang},
  \citenamefont {Pechal}, \citenamefont {Jiang}, \citenamefont {McKenna},
  \citenamefont {Witmer}, \citenamefont {Van~Laer},\ and\ \citenamefont
  {Safavi-Naeini}}]{arrango2019}%
  \BibitemOpen
  \bibfield  {author} {\bibinfo {author} {\bibfnamefont {P.}~\bibnamefont
  {Arrangoiz-Arriola}}, \bibinfo {author} {\bibfnamefont {E.~A.}\ \bibnamefont
  {Wollack}}, \bibinfo {author} {\bibfnamefont {Z.}~\bibnamefont {Wang}},
  \bibinfo {author} {\bibfnamefont {M.}~\bibnamefont {Pechal}}, \bibinfo
  {author} {\bibfnamefont {W.}~\bibnamefont {Jiang}}, \bibinfo {author}
  {\bibfnamefont {T.~P.}\ \bibnamefont {McKenna}}, \bibinfo {author}
  {\bibfnamefont {J.~D.}\ \bibnamefont {Witmer}}, \bibinfo {author}
  {\bibfnamefont {R.}~\bibnamefont {Van~Laer}},\ and\ \bibinfo {author}
  {\bibfnamefont {A.~H.}\ \bibnamefont {Safavi-Naeini}},\ }\bibfield  {title}
  {\bibinfo {title} {Resolving the energy levels of a nanomechanical
  oscillator},\ }\href {https://doi.org/10.1038/s41586-019-1386-x} {\bibfield
  {journal} {\bibinfo  {journal} {Nature}\ }\textbf {\bibinfo {volume} {571}},\
  \bibinfo {pages} {537} (\bibinfo {year} {2019})}\BibitemShut {NoStop}%
\bibitem [{\citenamefont {Qiao}\ \emph {et~al.}(2023)\citenamefont {Qiao},
  \citenamefont {Dumur}, \citenamefont {Andersson}, \citenamefont {Yan},
  \citenamefont {Chou}, \citenamefont {Grebel}, \citenamefont {Conner},
  \citenamefont {Joshi}, \citenamefont {Miller}, \citenamefont {Povey},
  \citenamefont {Wu},\ and\ \citenamefont {Cleland}}]{Qiao2023}%
  \BibitemOpen
  \bibfield  {author} {\bibinfo {author} {\bibfnamefont {H.}~\bibnamefont
  {Qiao}}, \bibinfo {author} {\bibfnamefont {{\'E}.}~\bibnamefont {Dumur}},
  \bibinfo {author} {\bibfnamefont {G.}~\bibnamefont {Andersson}}, \bibinfo
  {author} {\bibfnamefont {H.}~\bibnamefont {Yan}}, \bibinfo {author}
  {\bibfnamefont {M.-H.}\ \bibnamefont {Chou}}, \bibinfo {author}
  {\bibfnamefont {J.}~\bibnamefont {Grebel}}, \bibinfo {author} {\bibfnamefont
  {C.~R.}\ \bibnamefont {Conner}}, \bibinfo {author} {\bibfnamefont {Y.~J.}\
  \bibnamefont {Joshi}}, \bibinfo {author} {\bibfnamefont {J.~M.}\ \bibnamefont
  {Miller}}, \bibinfo {author} {\bibfnamefont {R.~G.}\ \bibnamefont {Povey}},
  \bibinfo {author} {\bibfnamefont {X.}~\bibnamefont {Wu}},\ and\ \bibinfo
  {author} {\bibfnamefont {A.~N.}\ \bibnamefont {Cleland}},\ }\bibfield
  {title} {\bibinfo {title} {Splitting phonons: Building a platform for linear
  mechanical quantum computing},\ }\href
  {https://doi.org/10.1126/science.adg8715} {\bibfield  {journal} {\bibinfo
  {journal} {Science}\ }\textbf {\bibinfo {volume} {380}},\ \bibinfo {pages}
  {1030} (\bibinfo {year} {2023})}\BibitemShut {NoStop}%
\bibitem [{\citenamefont {Chu}\ \emph {et~al.}(2018)\citenamefont {Chu},
  \citenamefont {Kharel}, \citenamefont {Yoon}, \citenamefont {Frunzio},
  \citenamefont {Rakich},\ and\ \citenamefont {Schoelkopf}}]{Chu:18}%
  \BibitemOpen
  \bibfield  {author} {\bibinfo {author} {\bibfnamefont {Y.}~\bibnamefont
  {Chu}}, \bibinfo {author} {\bibfnamefont {P.}~\bibnamefont {Kharel}},
  \bibinfo {author} {\bibfnamefont {T.}~\bibnamefont {Yoon}}, \bibinfo {author}
  {\bibfnamefont {L.}~\bibnamefont {Frunzio}}, \bibinfo {author} {\bibfnamefont
  {P.~T.}\ \bibnamefont {Rakich}},\ and\ \bibinfo {author} {\bibfnamefont
  {R.~J.}\ \bibnamefont {Schoelkopf}},\ }\bibfield  {title} {\bibinfo {title}
  {Creation and control of multi-phonon {Fock} states in a bulk acoustic-wave
  resonator},\ }\href {https://doi.org/10.1038/s41586-018-0717-7} {\bibfield
  {journal} {\bibinfo  {journal} {Nature}\ }\textbf {\bibinfo {volume} {563}},\
  \bibinfo {pages} {666} (\bibinfo {year} {2018})}\BibitemShut {NoStop}%
\bibitem [{\citenamefont {Bild}\ \emph {et~al.}(2023)\citenamefont {Bild},
  \citenamefont {Fadel}, \citenamefont {Yang}, \citenamefont {Von~L{\"u}pke},
  \citenamefont {Martin}, \citenamefont {Bruno},\ and\ \citenamefont
  {Chu}}]{Yiwen2023}%
  \BibitemOpen
  \bibfield  {author} {\bibinfo {author} {\bibfnamefont {M.}~\bibnamefont
  {Bild}}, \bibinfo {author} {\bibfnamefont {M.}~\bibnamefont {Fadel}},
  \bibinfo {author} {\bibfnamefont {Y.}~\bibnamefont {Yang}}, \bibinfo {author}
  {\bibfnamefont {U.}~\bibnamefont {Von~L{\"u}pke}}, \bibinfo {author}
  {\bibfnamefont {P.}~\bibnamefont {Martin}}, \bibinfo {author} {\bibfnamefont
  {A.}~\bibnamefont {Bruno}},\ and\ \bibinfo {author} {\bibfnamefont
  {Y.}~\bibnamefont {Chu}},\ }\bibfield  {title} {\bibinfo {title}
  {Schr{\"o}dinger cat states of a 16-microgram mechanical oscillator},\ }\href
  {https://www.science.org/doi/10.1126/science.adf7553} {\bibfield  {journal}
  {\bibinfo  {journal} {Science}\ }\textbf {\bibinfo {volume} {380}},\ \bibinfo
  {pages} {274} (\bibinfo {year} {2023})}\BibitemShut {NoStop}%
\bibitem [{\citenamefont {Yang}\ \emph {et~al.}(2024)\citenamefont {Yang},
  \citenamefont {Kladari{\'c}}, \citenamefont {Drimmer}, \citenamefont {von
  L{\"u}pke}, \citenamefont {Lenterman}, \citenamefont {Bus}, \citenamefont
  {Marti}, \citenamefont {Fadel},\ and\ \citenamefont {Chu}}]{Yiwen2024}%
  \BibitemOpen
  \bibfield  {author} {\bibinfo {author} {\bibfnamefont {Y.}~\bibnamefont
  {Yang}}, \bibinfo {author} {\bibfnamefont {I.}~\bibnamefont {Kladari{\'c}}},
  \bibinfo {author} {\bibfnamefont {M.}~\bibnamefont {Drimmer}}, \bibinfo
  {author} {\bibfnamefont {U.}~\bibnamefont {von L{\"u}pke}}, \bibinfo {author}
  {\bibfnamefont {D.}~\bibnamefont {Lenterman}}, \bibinfo {author}
  {\bibfnamefont {J.}~\bibnamefont {Bus}}, \bibinfo {author} {\bibfnamefont
  {S.}~\bibnamefont {Marti}}, \bibinfo {author} {\bibfnamefont
  {M.}~\bibnamefont {Fadel}},\ and\ \bibinfo {author} {\bibfnamefont
  {Y.}~\bibnamefont {Chu}},\ }\bibfield  {title} {\bibinfo {title} {A
  mechanical qubit},\ }\href
  {https://www.science.org/doi/10.1126/science.adr2464} {\bibfield  {journal}
  {\bibinfo  {journal} {Science}\ }\textbf {\bibinfo {volume} {386}},\ \bibinfo
  {pages} {783} (\bibinfo {year} {2024})}\BibitemShut {NoStop}%
\bibitem [{\citenamefont {Bienfait}\ \emph {et~al.}(2019)\citenamefont
  {Bienfait}, \citenamefont {Satzinger}, \citenamefont {Zhong}, \citenamefont
  {Chang}, \citenamefont {Chou}, \citenamefont {Conner}, \citenamefont {Dumur},
  \citenamefont {Grebel}, \citenamefont {Peairs}, \citenamefont {Povey} \emph
  {et~al.}}]{Cleland1}%
  \BibitemOpen
  \bibfield  {author} {\bibinfo {author} {\bibfnamefont {A.}~\bibnamefont
  {Bienfait}}, \bibinfo {author} {\bibfnamefont {K.~J.}\ \bibnamefont
  {Satzinger}}, \bibinfo {author} {\bibfnamefont {Y.}~\bibnamefont {Zhong}},
  \bibinfo {author} {\bibfnamefont {H.-S.}\ \bibnamefont {Chang}}, \bibinfo
  {author} {\bibfnamefont {M.-H.}\ \bibnamefont {Chou}}, \bibinfo {author}
  {\bibfnamefont {C.~R.}\ \bibnamefont {Conner}}, \bibinfo {author}
  {\bibfnamefont {{\'E}.}~\bibnamefont {Dumur}}, \bibinfo {author}
  {\bibfnamefont {J.}~\bibnamefont {Grebel}}, \bibinfo {author} {\bibfnamefont
  {G.~A.}\ \bibnamefont {Peairs}}, \bibinfo {author} {\bibfnamefont {R.~G.}\
  \bibnamefont {Povey}}, \emph {et~al.},\ }\bibfield  {title} {\bibinfo {title}
  {Phonon-mediated quantum state transfer and remote qubit entanglement},\
  }\href {https://www.science.org/doi/10.1126/science.aaw8415} {\bibfield
  {journal} {\bibinfo  {journal} {Science}\ }\textbf {\bibinfo {volume}
  {364}},\ \bibinfo {pages} {368} (\bibinfo {year} {2019})}\BibitemShut
  {NoStop}%
\bibitem [{\citenamefont {Chou}\ \emph {et~al.}(2025)\citenamefont {Chou},
  \citenamefont {Qiao}, \citenamefont {Yan}, \citenamefont {Andersson},
  \citenamefont {Conner}, \citenamefont {Grebel}, \citenamefont {Joshi},
  \citenamefont {Miller}, \citenamefont {Povey}, \citenamefont {Wu} \emph
  {et~al.}}]{Cleland2}%
  \BibitemOpen
  \bibfield  {author} {\bibinfo {author} {\bibfnamefont {M.-H.}\ \bibnamefont
  {Chou}}, \bibinfo {author} {\bibfnamefont {H.}~\bibnamefont {Qiao}}, \bibinfo
  {author} {\bibfnamefont {H.}~\bibnamefont {Yan}}, \bibinfo {author}
  {\bibfnamefont {G.}~\bibnamefont {Andersson}}, \bibinfo {author}
  {\bibfnamefont {C.~R.}\ \bibnamefont {Conner}}, \bibinfo {author}
  {\bibfnamefont {J.}~\bibnamefont {Grebel}}, \bibinfo {author} {\bibfnamefont
  {Y.~J.}\ \bibnamefont {Joshi}}, \bibinfo {author} {\bibfnamefont {J.~M.}\
  \bibnamefont {Miller}}, \bibinfo {author} {\bibfnamefont {R.~G.}\
  \bibnamefont {Povey}}, \bibinfo {author} {\bibfnamefont {X.}~\bibnamefont
  {Wu}}, \emph {et~al.},\ }\bibfield  {title} {\bibinfo {title} {Deterministic
  multi-phonon entanglement between two mechanical resonators on separate
  substrates},\ }\href {https://doi.org/10.1038/s41467-025-56454-0} {\bibfield
  {journal} {\bibinfo  {journal} {Nature Communications}\ }\textbf {\bibinfo
  {volume} {16}},\ \bibinfo {pages} {1450} (\bibinfo {year}
  {2025})}\BibitemShut {NoStop}%
\bibitem [{\citenamefont {von Lüpke}\ \emph {et~al.}(2024)\citenamefont {von
  Lüpke}, \citenamefont {Rodrigues}, \citenamefont {Yang}, \citenamefont
  {Fadel},\ and\ \citenamefont {Chu}}]{von_Lupke:24}%
  \BibitemOpen
  \bibfield  {author} {\bibinfo {author} {\bibfnamefont {U.}~\bibnamefont {von
  Lüpke}}, \bibinfo {author} {\bibfnamefont {I.~C.}\ \bibnamefont
  {Rodrigues}}, \bibinfo {author} {\bibfnamefont {Y.}~\bibnamefont {Yang}},
  \bibinfo {author} {\bibfnamefont {M.}~\bibnamefont {Fadel}},\ and\ \bibinfo
  {author} {\bibfnamefont {Y.}~\bibnamefont {Chu}},\ }\bibfield  {title}
  {\bibinfo {title} {Engineering multimode interactions in circuit quantum
  acoustodynamics},\ }\href {https://doi.org/10.1038/s41567-023-02377-w}
  {\bibfield  {journal} {\bibinfo  {journal} {Nature Physics}\ }\textbf
  {\bibinfo {volume} {20}},\ \bibinfo {pages} {564} (\bibinfo {year}
  {2024})}\BibitemShut {NoStop}%
\bibitem [{\citenamefont {Kervinen}\ \emph {et~al.}(2019)\citenamefont
  {Kervinen}, \citenamefont {Ram\'{\i}rez-Mu\~noz}, \citenamefont {V\"alimaa},\
  and\ \citenamefont {Sillanp\"a\"a}}]{Kervinen19}%
  \BibitemOpen
  \bibfield  {author} {\bibinfo {author} {\bibfnamefont {M.}~\bibnamefont
  {Kervinen}}, \bibinfo {author} {\bibfnamefont {J.~E.}\ \bibnamefont
  {Ram\'{\i}rez-Mu\~noz}}, \bibinfo {author} {\bibfnamefont {A.}~\bibnamefont
  {V\"alimaa}},\ and\ \bibinfo {author} {\bibfnamefont {M.~A.}\ \bibnamefont
  {Sillanp\"a\"a}},\ }\bibfield  {title} {\bibinfo {title}
  {Landau-zener-st\"uckelberg interference in a multimode electromechanical
  system in the quantum regime},\ }\href
  {https://doi.org/10.1103/PhysRevLett.123.240401} {\bibfield  {journal}
  {\bibinfo  {journal} {Phys. Rev. Lett.}\ }\textbf {\bibinfo {volume} {123}},\
  \bibinfo {pages} {240401} (\bibinfo {year} {2019})}\BibitemShut {NoStop}%
\bibitem [{\citenamefont {Mirhosseini}\ \emph {et~al.}(2020)\citenamefont
  {Mirhosseini}, \citenamefont {Sipahigil}, \citenamefont {Kalaee},\ and\
  \citenamefont {Painter}}]{Mirhosseini20}%
  \BibitemOpen
  \bibfield  {author} {\bibinfo {author} {\bibfnamefont {M.}~\bibnamefont
  {Mirhosseini}}, \bibinfo {author} {\bibfnamefont {A.}~\bibnamefont
  {Sipahigil}}, \bibinfo {author} {\bibfnamefont {M.}~\bibnamefont {Kalaee}},\
  and\ \bibinfo {author} {\bibfnamefont {O.}~\bibnamefont {Painter}},\
  }\bibfield  {title} {\bibinfo {title} {Superconducting qubit to optical
  photon transduction},\ }\href
  {https://doi.org/https://doi.org/10.1038/s41586-020-3038-6} {\bibfield
  {journal} {\bibinfo  {journal} {Nature}\ }\textbf {\bibinfo {volume} {588}},\
  \bibinfo {pages} {599} (\bibinfo {year} {2020})}\BibitemShut {NoStop}%
\bibitem [{\citenamefont {Kitzman}\ \emph {et~al.}(2023)\citenamefont
  {Kitzman}, \citenamefont {Lane}, \citenamefont {Undershute}, \citenamefont
  {Harrington}, \citenamefont {Beysengulov}, \citenamefont {Mikolas},
  \citenamefont {Murch},\ and\ \citenamefont {Pollanen}}]{Yohannes2023}%
  \BibitemOpen
  \bibfield  {author} {\bibinfo {author} {\bibfnamefont {J.~M.}\ \bibnamefont
  {Kitzman}}, \bibinfo {author} {\bibfnamefont {J.~R.}\ \bibnamefont {Lane}},
  \bibinfo {author} {\bibfnamefont {C.}~\bibnamefont {Undershute}}, \bibinfo
  {author} {\bibfnamefont {P.~M.}\ \bibnamefont {Harrington}}, \bibinfo
  {author} {\bibfnamefont {N.~R.}\ \bibnamefont {Beysengulov}}, \bibinfo
  {author} {\bibfnamefont {C.~A.}\ \bibnamefont {Mikolas}}, \bibinfo {author}
  {\bibfnamefont {K.~W.}\ \bibnamefont {Murch}},\ and\ \bibinfo {author}
  {\bibfnamefont {J.}~\bibnamefont {Pollanen}},\ }\bibfield  {title} {\bibinfo
  {title} {Phononic bath engineering of a superconducting qubit},\ }\href
  {https://doi.org/10.1038/s41467-023-39682-0} {\bibfield  {journal} {\bibinfo
  {journal} {Nature Communications}\ }\textbf {\bibinfo {volume} {14}},\
  \bibinfo {pages} {3910} (\bibinfo {year} {2023})}\BibitemShut {NoStop}%
\bibitem [{\citenamefont {Bozkurt}\ \emph {et~al.}(2025)\citenamefont
  {Bozkurt}, \citenamefont {Golami}, \citenamefont {Yu}, \citenamefont {Tian},\
  and\ \citenamefont {Mirhosseini}}]{Bozkurt25}%
  \BibitemOpen
  \bibfield  {author} {\bibinfo {author} {\bibfnamefont {A.~B.}\ \bibnamefont
  {Bozkurt}}, \bibinfo {author} {\bibfnamefont {O.}~\bibnamefont {Golami}},
  \bibinfo {author} {\bibfnamefont {Y.}~\bibnamefont {Yu}}, \bibinfo {author}
  {\bibfnamefont {H.}~\bibnamefont {Tian}},\ and\ \bibinfo {author}
  {\bibfnamefont {M.}~\bibnamefont {Mirhosseini}},\ }\bibfield  {title}
  {\bibinfo {title} {A mechanical quantum memory for microwave photons},\
  }\href {https://doi.org/https://doi.org/10.1038/s41567-025-02975-w}
  {\bibfield  {journal} {\bibinfo  {journal} {Nature Physics}\ }\textbf
  {\bibinfo {volume} {21}},\ \bibinfo {pages} {1745} (\bibinfo {year}
  {2025})}\BibitemShut {NoStop}%
\bibitem [{\citenamefont {Chen}\ \emph {et~al.}(2025)\citenamefont {Chen},
  \citenamefont {He}, \citenamefont {Gao}, \citenamefont {Liu}, \citenamefont
  {Niu}, \citenamefont {Liu}, \citenamefont {Peng}, \citenamefont {Wang},\ and\
  \citenamefont {Lin}}]{Lin2025}%
  \BibitemOpen
  \bibfield  {author} {\bibinfo {author} {\bibfnamefont {J.}~\bibnamefont
  {Chen}}, \bibinfo {author} {\bibfnamefont {X.}~\bibnamefont {He}}, \bibinfo
  {author} {\bibfnamefont {W.}~\bibnamefont {Gao}}, \bibinfo {author}
  {\bibfnamefont {X.}~\bibnamefont {Liu}}, \bibinfo {author} {\bibfnamefont
  {Z.}~\bibnamefont {Niu}}, \bibinfo {author} {\bibfnamefont {K.}~\bibnamefont
  {Liu}}, \bibinfo {author} {\bibfnamefont {W.}~\bibnamefont {Peng}}, \bibinfo
  {author} {\bibfnamefont {Z.}~\bibnamefont {Wang}},\ and\ \bibinfo {author}
  {\bibfnamefont {Z.-R.}\ \bibnamefont {Lin}},\ }\bibfield  {title} {\bibinfo
  {title} {On-chip integration and strong coupling between scaln thin-film
  surface acoustic wave resonators and superconducting qubits},\ }\href
  {https://doi.org/10.1063/5.0258397} {\bibfield  {journal} {\bibinfo
  {journal} {Applied Physics Letters}\ }\textbf {\bibinfo {volume} {126}},\
  \bibinfo {pages} {144001} (\bibinfo {year} {2025})}\BibitemShut {NoStop}%
\bibitem [{\citenamefont {Wang}\ \emph {et~al.}(2025)\citenamefont {Wang},
  \citenamefont {Xiao}, \citenamefont {Zhang}, \citenamefont {Zeng},
  \citenamefont {Hua}, \citenamefont {Ma}, \citenamefont {Huang}, \citenamefont
  {Xu}, \citenamefont {Wang}, \citenamefont {Xue}, \citenamefont {Yu},
  \citenamefont {Xu}, \citenamefont {Zou},\ and\ \citenamefont {Sun}}]{arXiv1}%
  \BibitemOpen
  \bibfield  {author} {\bibinfo {author} {\bibfnamefont {W.}~\bibnamefont
  {Wang}}, \bibinfo {author} {\bibfnamefont {L.}~\bibnamefont {Xiao}}, \bibinfo
  {author} {\bibfnamefont {B.}~\bibnamefont {Zhang}}, \bibinfo {author}
  {\bibfnamefont {Y.}~\bibnamefont {Zeng}}, \bibinfo {author} {\bibfnamefont
  {Z.}~\bibnamefont {Hua}}, \bibinfo {author} {\bibfnamefont {C.}~\bibnamefont
  {Ma}}, \bibinfo {author} {\bibfnamefont {H.}~\bibnamefont {Huang}}, \bibinfo
  {author} {\bibfnamefont {Y.}~\bibnamefont {Xu}}, \bibinfo {author}
  {\bibfnamefont {J.-Q.}\ \bibnamefont {Wang}}, \bibinfo {author}
  {\bibfnamefont {G.}~\bibnamefont {Xue}}, \bibinfo {author} {\bibfnamefont
  {H.}~\bibnamefont {Yu}}, \bibinfo {author} {\bibfnamefont {X.-B.}\
  \bibnamefont {Xu}}, \bibinfo {author} {\bibfnamefont {C.-L.}\ \bibnamefont
  {Zou}},\ and\ \bibinfo {author} {\bibfnamefont {L.}~\bibnamefont {Sun}},\
  }\bibfield  {title} {\bibinfo {title} {Circuit quantum acoustodynamics in a
  scalable phononic integrated circuit architecture},\ }\href@noop {}
  {\bibfield  {journal} {\bibinfo  {journal} {arXiv:2512.04953}\ } (\bibinfo
  {year} {2025})}\BibitemShut {NoStop}%
\bibitem [{\citenamefont {Li}\ \emph {et~al.}(2025)\citenamefont {Li},
  \citenamefont {Zhao}, \citenamefont {Chen}, \citenamefont {Liang},
  \citenamefont {Liu}, \citenamefont {Deng}, \citenamefont {Yuan},
  \citenamefont {Song}, \citenamefont {Liu}, \citenamefont {Li}, \citenamefont
  {Shi}, \citenamefont {Zhang}, \citenamefont {Han}, \citenamefont {Guo},
  \citenamefont {Guo}, \citenamefont {Song}, \citenamefont {Zhao},
  \citenamefont {Zhang}, \citenamefont {Song}, \citenamefont {Xu},
  \citenamefont {Fan}, \citenamefont {Liu}, \citenamefont {Peng}, \citenamefont
  {Xiang},\ and\ \citenamefont {Zheng}}]{arXiv2}%
  \BibitemOpen
  \bibfield  {author} {\bibinfo {author} {\bibfnamefont {X.}~\bibnamefont {Li},
  \bibfnamefont {Li~annd~Ruan}}, \bibinfo {author} {\bibfnamefont {S.-L.}\
  \bibnamefont {Zhao}}, \bibinfo {author} {\bibfnamefont {B.-J.}\ \bibnamefont
  {Chen}}, \bibinfo {author} {\bibfnamefont {G.-H.}\ \bibnamefont {Liang}},
  \bibinfo {author} {\bibfnamefont {Y.}~\bibnamefont {Liu}}, \bibinfo {author}
  {\bibfnamefont {C.-L.}\ \bibnamefont {Deng}}, \bibinfo {author}
  {\bibfnamefont {W.-P.}\ \bibnamefont {Yuan}}, \bibinfo {author}
  {\bibfnamefont {J.-C.}\ \bibnamefont {Song}}, \bibinfo {author}
  {\bibfnamefont {Z.-H.}\ \bibnamefont {Liu}}, \bibinfo {author} {\bibfnamefont
  {T.-M.}\ \bibnamefont {Li}}, \bibinfo {author} {\bibfnamefont {Y.-H.}\
  \bibnamefont {Shi}}, \bibinfo {author} {\bibfnamefont {H.}~\bibnamefont
  {Zhang}}, \bibinfo {author} {\bibfnamefont {M.}~\bibnamefont {Han}}, \bibinfo
  {author} {\bibfnamefont {J.-M.}\ \bibnamefont {Guo}}, \bibinfo {author}
  {\bibfnamefont {X.-Y.}\ \bibnamefont {Guo}}, \bibinfo {author} {\bibfnamefont
  {X.}~\bibnamefont {Song}}, \bibinfo {author} {\bibfnamefont {Q.}~\bibnamefont
  {Zhao}}, \bibinfo {author} {\bibfnamefont {J.}~\bibnamefont {Zhang}},
  \bibinfo {author} {\bibfnamefont {P.}~\bibnamefont {Song}}, \bibinfo {author}
  {\bibfnamefont {K.}~\bibnamefont {Xu}}, \bibinfo {author} {\bibfnamefont
  {H.}~\bibnamefont {Fan}}, \bibinfo {author} {\bibfnamefont {Y.-X.}\
  \bibnamefont {Liu}}, \bibinfo {author} {\bibfnamefont {Z.}~\bibnamefont
  {Peng}}, \bibinfo {author} {\bibfnamefont {Z.}~\bibnamefont {Xiang}},\ and\
  \bibinfo {author} {\bibfnamefont {D.}~\bibnamefont {Zheng}},\ }\bibfield
  {title} {\bibinfo {title} {Quantum acoustics with superconducting qubits in
  the multimode transition-coupling regime},\ }\href@noop {} {\bibfield
  {journal} {\bibinfo  {journal} {arXiv: 2505.05127}\ } (\bibinfo {year}
  {2025})}\BibitemShut {NoStop}%
\bibitem [{\citenamefont {Gokhale}\ \emph {et~al.}(2020)\citenamefont
  {Gokhale}, \citenamefont {Downey}, \citenamefont {Katzer}, \citenamefont
  {Nepal}, \citenamefont {Lang}, \citenamefont {Stroud},\ and\ \citenamefont
  {Meyer}}]{Epitaxy20}%
  \BibitemOpen
  \bibfield  {author} {\bibinfo {author} {\bibfnamefont {V.~J.}\ \bibnamefont
  {Gokhale}}, \bibinfo {author} {\bibfnamefont {B.~P.}\ \bibnamefont {Downey}},
  \bibinfo {author} {\bibfnamefont {D.~S.}\ \bibnamefont {Katzer}}, \bibinfo
  {author} {\bibfnamefont {N.}~\bibnamefont {Nepal}}, \bibinfo {author}
  {\bibfnamefont {A.~C.}\ \bibnamefont {Lang}}, \bibinfo {author}
  {\bibfnamefont {R.~M.}\ \bibnamefont {Stroud}},\ and\ \bibinfo {author}
  {\bibfnamefont {D.~J.}\ \bibnamefont {Meyer}},\ }\bibfield  {title} {\bibinfo
  {title} {Epitaxial bulk acoustic wave resonators as highly coherent
  multi-phonon sources for quantum acoustodynamics},\ }\href
  {https://doi.org/https://doi.org/10.1038/s41467-020-15472-w} {\bibfield
  {journal} {\bibinfo  {journal} {Nature communications}\ }\textbf {\bibinfo
  {volume} {11}},\ \bibinfo {pages} {2314} (\bibinfo {year}
  {2020})}\BibitemShut {NoStop}%
\bibitem [{SM()}]{SM}%
  \BibitemOpen
  \href@noop {} {}\bibinfo {note} {See Supplemental Material at
  [{\color{blue}URL will be inserted by publisher}] for further details on the
  theoretical model of the clustered HBAR, in addition to analysis of the
  collective behavior and entanglement creation. We also present details on the
  device fabrication, experimental setup, qubit control and readout
  corrections, device parameters, qubit spectroscopy and further experimental
  data for phononic Rabi oscillations. The Supplemental Material also includes
  Refs.~\cite{Crump:23,Chu:17,Chu:18,von_Lupke:24,johansson2012qutip}}\BibitemShut
  {NoStop}%
\bibitem [{\citenamefont {Shkarin}\ \emph {et~al.}(2014)\citenamefont
  {Shkarin}, \citenamefont {Flowers-Jacobs}, \citenamefont {Hoch},
  \citenamefont {Kashkanova}, \citenamefont {Deutsch}, \citenamefont
  {Reichel},\ and\ \citenamefont {Harris}}]{Shkarin:14}%
  \BibitemOpen
  \bibfield  {author} {\bibinfo {author} {\bibfnamefont {A.~B.}\ \bibnamefont
  {Shkarin}}, \bibinfo {author} {\bibfnamefont {N.~E.}\ \bibnamefont
  {Flowers-Jacobs}}, \bibinfo {author} {\bibfnamefont {S.~W.}\ \bibnamefont
  {Hoch}}, \bibinfo {author} {\bibfnamefont {A.~D.}\ \bibnamefont
  {Kashkanova}}, \bibinfo {author} {\bibfnamefont {C.}~\bibnamefont {Deutsch}},
  \bibinfo {author} {\bibfnamefont {J.}~\bibnamefont {Reichel}},\ and\ \bibinfo
  {author} {\bibfnamefont {J.~G.~E.}\ \bibnamefont {Harris}},\ }\bibfield
  {title} {\bibinfo {title} {Optically mediated hybridization between two
  mechanical modes},\ }\href {https://doi.org/10.1103/PhysRevLett.112.013602}
  {\bibfield  {journal} {\bibinfo  {journal} {Phys. Rev. Lett.}\ }\textbf
  {\bibinfo {volume} {112}},\ \bibinfo {pages} {013602} (\bibinfo {year}
  {2014})}\BibitemShut {NoStop}%
\bibitem [{\citenamefont {Kharel}\ \emph {et~al.}(2022)\citenamefont {Kharel},
  \citenamefont {Chu}, \citenamefont {Mason}, \citenamefont {Kittlaus},
  \citenamefont {Otterstrom}, \citenamefont {Gertler},\ and\ \citenamefont
  {Rakich}}]{Kharel:22}%
  \BibitemOpen
  \bibfield  {author} {\bibinfo {author} {\bibfnamefont {P.}~\bibnamefont
  {Kharel}}, \bibinfo {author} {\bibfnamefont {Y.}~\bibnamefont {Chu}},
  \bibinfo {author} {\bibfnamefont {D.}~\bibnamefont {Mason}}, \bibinfo
  {author} {\bibfnamefont {E.~A.}\ \bibnamefont {Kittlaus}}, \bibinfo {author}
  {\bibfnamefont {N.~T.}\ \bibnamefont {Otterstrom}}, \bibinfo {author}
  {\bibfnamefont {S.}~\bibnamefont {Gertler}},\ and\ \bibinfo {author}
  {\bibfnamefont {P.~T.}\ \bibnamefont {Rakich}},\ }\bibfield  {title}
  {\bibinfo {title} {Multimode strong coupling in cavity optomechanics},\
  }\href {https://doi.org/10.1103/PhysRevApplied.18.024054} {\bibfield
  {journal} {\bibinfo  {journal} {Phys. Rev. Appl.}\ }\textbf {\bibinfo
  {volume} {18}},\ \bibinfo {pages} {024054} (\bibinfo {year}
  {2022})}\BibitemShut {NoStop}%
\bibitem [{\citenamefont {Barzanjeh}\ \emph {et~al.}(2022)\citenamefont
  {Barzanjeh}, \citenamefont {Xuereb}, \citenamefont {Gr{\"o}blacher},
  \citenamefont {Paternostro}, \citenamefont {Regal},\ and\ \citenamefont
  {Weig}}]{Barzanjeh:22}%
  \BibitemOpen
  \bibfield  {author} {\bibinfo {author} {\bibfnamefont {S.}~\bibnamefont
  {Barzanjeh}}, \bibinfo {author} {\bibfnamefont {A.}~\bibnamefont {Xuereb}},
  \bibinfo {author} {\bibfnamefont {S.}~\bibnamefont {Gr{\"o}blacher}},
  \bibinfo {author} {\bibfnamefont {M.}~\bibnamefont {Paternostro}}, \bibinfo
  {author} {\bibfnamefont {C.~A.}\ \bibnamefont {Regal}},\ and\ \bibinfo
  {author} {\bibfnamefont {E.~M.}\ \bibnamefont {Weig}},\ }\bibfield  {title}
  {\bibinfo {title} {Optomechanics for quantum technologies},\ }\href
  {https://doi.org/https://doi.org/10.1038/s41567-021-01402-0} {\bibfield
  {journal} {\bibinfo  {journal} {Nature Physics}\ }\textbf {\bibinfo {volume}
  {18}},\ \bibinfo {pages} {15} (\bibinfo {year} {2022})}\BibitemShut {NoStop}%
\bibitem [{\citenamefont {Crump}\ \emph {et~al.}(2023)\citenamefont {Crump},
  \citenamefont {Välimaa},\ and\ \citenamefont {Sillanpää}}]{Crump:23}%
  \BibitemOpen
  \bibfield  {author} {\bibinfo {author} {\bibfnamefont {W.}~\bibnamefont
  {Crump}}, \bibinfo {author} {\bibfnamefont {A.}~\bibnamefont {Välimaa}},\
  and\ \bibinfo {author} {\bibfnamefont {M.~A.}\ \bibnamefont {Sillanpää}},\
  }\bibfield  {title} {\bibinfo {title} {Coupling high-overtone bulk acoustic
  wave resonators via superconducting qubits},\ }\href
  {https://doi.org/10.1063/5.0166924} {\bibfield  {journal} {\bibinfo
  {journal} {Applied Physics Letters}\ }\textbf {\bibinfo {volume} {123}},\
  \bibinfo {pages} {134004} (\bibinfo {year} {2023})}\BibitemShut {NoStop}%
\bibitem [{\citenamefont {Pennetta}\ \emph
  {et~al.}(2022{\natexlab{a}})\citenamefont {Pennetta}, \citenamefont {Blaha},
  \citenamefont {Johnson}, \citenamefont {Lechner}, \citenamefont
  {Schneeweiss}, \citenamefont {Volz},\ and\ \citenamefont
  {Rauschenbeutel}}]{Arno2022}%
  \BibitemOpen
  \bibfield  {author} {\bibinfo {author} {\bibfnamefont {R.}~\bibnamefont
  {Pennetta}}, \bibinfo {author} {\bibfnamefont {M.}~\bibnamefont {Blaha}},
  \bibinfo {author} {\bibfnamefont {A.}~\bibnamefont {Johnson}}, \bibinfo
  {author} {\bibfnamefont {D.}~\bibnamefont {Lechner}}, \bibinfo {author}
  {\bibfnamefont {P.}~\bibnamefont {Schneeweiss}}, \bibinfo {author}
  {\bibfnamefont {J.}~\bibnamefont {Volz}},\ and\ \bibinfo {author}
  {\bibfnamefont {A.}~\bibnamefont {Rauschenbeutel}},\ }\bibfield  {title}
  {\bibinfo {title} {Collective radiative dynamics of an ensemble of cold atoms
  coupled to an optical waveguide},\ }\href
  {https://doi.org/10.1103/PhysRevLett.128.073601} {\bibfield  {journal}
  {\bibinfo  {journal} {Phys. Rev. Lett.}\ }\textbf {\bibinfo {volume} {128}},\
  \bibinfo {pages} {073601} (\bibinfo {year} {2022}{\natexlab{a}})}\BibitemShut
  {NoStop}%
\bibitem [{\citenamefont {Pennetta}\ \emph
  {et~al.}(2022{\natexlab{b}})\citenamefont {Pennetta}, \citenamefont
  {Lechner}, \citenamefont {Blaha}, \citenamefont {Rauschenbeutel},
  \citenamefont {Schneeweiss},\ and\ \citenamefont {Volz}}]{Pennetta2022}%
  \BibitemOpen
  \bibfield  {author} {\bibinfo {author} {\bibfnamefont {R.}~\bibnamefont
  {Pennetta}}, \bibinfo {author} {\bibfnamefont {D.}~\bibnamefont {Lechner}},
  \bibinfo {author} {\bibfnamefont {M.}~\bibnamefont {Blaha}}, \bibinfo
  {author} {\bibfnamefont {A.}~\bibnamefont {Rauschenbeutel}}, \bibinfo
  {author} {\bibfnamefont {P.}~\bibnamefont {Schneeweiss}},\ and\ \bibinfo
  {author} {\bibfnamefont {J.}~\bibnamefont {Volz}},\ }\bibfield  {title}
  {\bibinfo {title} {Observation of coherent coupling between super- and
  subradiant states of an ensemble of cold atoms collectively coupled to a
  single propagating optical mode},\ }\href
  {https://doi.org/10.1103/PhysRevLett.128.203601} {\bibfield  {journal}
  {\bibinfo  {journal} {Phys. Rev. Lett.}\ }\textbf {\bibinfo {volume} {128}},\
  \bibinfo {pages} {203601} (\bibinfo {year} {2022}{\natexlab{b}})}\BibitemShut
  {NoStop}%
\bibitem [{\citenamefont {Scully}(2009)}]{Scully:09}%
  \BibitemOpen
  \bibfield  {author} {\bibinfo {author} {\bibfnamefont {M.~O.}\ \bibnamefont
  {Scully}},\ }\bibfield  {title} {\bibinfo {title} {Collective lamb shift in
  single photon dicke superradiance},\ }\href
  {https://doi.org/10.1103/PhysRevLett.102.143601} {\bibfield  {journal}
  {\bibinfo  {journal} {Phys. Rev. Lett.}\ }\textbf {\bibinfo {volume} {102}},\
  \bibinfo {pages} {143601} (\bibinfo {year} {2009})}\BibitemShut {NoStop}%
\bibitem [{\citenamefont {Guerin}\ \emph {et~al.}(2016)\citenamefont {Guerin},
  \citenamefont {Ara\'ujo},\ and\ \citenamefont {Kaiser}}]{Guerin:16}%
  \BibitemOpen
  \bibfield  {author} {\bibinfo {author} {\bibfnamefont {W.}~\bibnamefont
  {Guerin}}, \bibinfo {author} {\bibfnamefont {M.~O.}\ \bibnamefont
  {Ara\'ujo}},\ and\ \bibinfo {author} {\bibfnamefont {R.}~\bibnamefont
  {Kaiser}},\ }\bibfield  {title} {\bibinfo {title} {Subradiance in a large
  cloud of cold atoms},\ }\href
  {https://doi.org/10.1103/PhysRevLett.116.083601} {\bibfield  {journal}
  {\bibinfo  {journal} {Phys. Rev. Lett.}\ }\textbf {\bibinfo {volume} {116}},\
  \bibinfo {pages} {083601} (\bibinfo {year} {2016})}\BibitemShut {NoStop}%
\bibitem [{\citenamefont {Rui}\ \emph {et~al.}(2020)\citenamefont {Rui},
  \citenamefont {Wei}, \citenamefont {Rubio-Abadal}, \citenamefont {Hollerith},
  \citenamefont {Zeiher}, \citenamefont {Stamper-Kurn}, \citenamefont {Gross},\
  and\ \citenamefont {Bloch}}]{Rui:20}%
  \BibitemOpen
  \bibfield  {author} {\bibinfo {author} {\bibfnamefont {J.}~\bibnamefont
  {Rui}}, \bibinfo {author} {\bibfnamefont {D.}~\bibnamefont {Wei}}, \bibinfo
  {author} {\bibfnamefont {A.}~\bibnamefont {Rubio-Abadal}}, \bibinfo {author}
  {\bibfnamefont {S.}~\bibnamefont {Hollerith}}, \bibinfo {author}
  {\bibfnamefont {J.}~\bibnamefont {Zeiher}}, \bibinfo {author} {\bibfnamefont
  {D.~M.}\ \bibnamefont {Stamper-Kurn}}, \bibinfo {author} {\bibfnamefont
  {C.}~\bibnamefont {Gross}},\ and\ \bibinfo {author} {\bibfnamefont
  {I.}~\bibnamefont {Bloch}},\ }\bibfield  {title} {\bibinfo {title} {A
  subradiant optical mirror formed by a single structured atomic layer},\
  }\href {https://doi.org/https://doi.org/10.1038/s41586-020-2463-x} {\bibfield
   {journal} {\bibinfo  {journal} {Nature}\ }\textbf {\bibinfo {volume}
  {583}},\ \bibinfo {pages} {369} (\bibinfo {year} {2020})}\BibitemShut
  {NoStop}%
\bibitem [{\citenamefont {He}\ \emph {et~al.}(2020{\natexlab{a}})\citenamefont
  {He}, \citenamefont {Ji}, \citenamefont {Wang}, \citenamefont {Qiu},
  \citenamefont {Zhao}, \citenamefont {Ma}, \citenamefont {Huang},
  \citenamefont {Wu},\ and\ \citenamefont {Chang}}]{He:20}%
  \BibitemOpen
  \bibfield  {author} {\bibinfo {author} {\bibfnamefont {Y.}~\bibnamefont
  {He}}, \bibinfo {author} {\bibfnamefont {L.}~\bibnamefont {Ji}}, \bibinfo
  {author} {\bibfnamefont {Y.}~\bibnamefont {Wang}}, \bibinfo {author}
  {\bibfnamefont {L.}~\bibnamefont {Qiu}}, \bibinfo {author} {\bibfnamefont
  {J.}~\bibnamefont {Zhao}}, \bibinfo {author} {\bibfnamefont {Y.}~\bibnamefont
  {Ma}}, \bibinfo {author} {\bibfnamefont {X.}~\bibnamefont {Huang}}, \bibinfo
  {author} {\bibfnamefont {S.}~\bibnamefont {Wu}},\ and\ \bibinfo {author}
  {\bibfnamefont {D.~E.}\ \bibnamefont {Chang}},\ }\bibfield  {title} {\bibinfo
  {title} {Geometric control of collective spontaneous emission},\ }\href
  {https://doi.org/10.1103/PhysRevLett.125.213602} {\bibfield  {journal}
  {\bibinfo  {journal} {Phys. Rev. Lett.}\ }\textbf {\bibinfo {volume} {125}},\
  \bibinfo {pages} {213602} (\bibinfo {year} {2020}{\natexlab{a}})}\BibitemShut
  {NoStop}%
\bibitem [{\citenamefont {Liedl}\ \emph {et~al.}(2023)\citenamefont {Liedl},
  \citenamefont {Pucher}, \citenamefont {Tebbenjohanns}, \citenamefont
  {Schneeweiss},\ and\ \citenamefont {Rauschenbeutel}}]{Liedl:23}%
  \BibitemOpen
  \bibfield  {author} {\bibinfo {author} {\bibfnamefont {C.}~\bibnamefont
  {Liedl}}, \bibinfo {author} {\bibfnamefont {S.}~\bibnamefont {Pucher}},
  \bibinfo {author} {\bibfnamefont {F.}~\bibnamefont {Tebbenjohanns}}, \bibinfo
  {author} {\bibfnamefont {P.}~\bibnamefont {Schneeweiss}},\ and\ \bibinfo
  {author} {\bibfnamefont {A.}~\bibnamefont {Rauschenbeutel}},\ }\bibfield
  {title} {\bibinfo {title} {Collective radiation of a cascaded quantum system:
  From timed dicke states to inverted ensembles},\ }\href
  {https://doi.org/10.1103/PhysRevLett.130.163602} {\bibfield  {journal}
  {\bibinfo  {journal} {Phys. Rev. Lett.}\ }\textbf {\bibinfo {volume} {130}},\
  \bibinfo {pages} {163602} (\bibinfo {year} {2023})}\BibitemShut {NoStop}%
\bibitem [{\citenamefont {Scully}\ \emph {et~al.}(2006)\citenamefont {Scully},
  \citenamefont {Fry}, \citenamefont {Ooi},\ and\ \citenamefont
  {W\'odkiewicz}}]{Scully:06}%
  \BibitemOpen
  \bibfield  {author} {\bibinfo {author} {\bibfnamefont {M.~O.}\ \bibnamefont
  {Scully}}, \bibinfo {author} {\bibfnamefont {E.~S.}\ \bibnamefont {Fry}},
  \bibinfo {author} {\bibfnamefont {C.~H.~R.}\ \bibnamefont {Ooi}},\ and\
  \bibinfo {author} {\bibfnamefont {K.}~\bibnamefont {W\'odkiewicz}},\
  }\bibfield  {title} {\bibinfo {title} {Directed spontaneous emission from an
  extended ensemble of $n$ atoms: Timing is everything},\ }\href
  {https://doi.org/10.1103/PhysRevLett.96.010501} {\bibfield  {journal}
  {\bibinfo  {journal} {Phys. Rev. Lett.}\ }\textbf {\bibinfo {volume} {96}},\
  \bibinfo {pages} {010501} (\bibinfo {year} {2006})}\BibitemShut {NoStop}%
\bibitem [{\citenamefont {He}\ \emph {et~al.}(2020{\natexlab{b}})\citenamefont
  {He}, \citenamefont {Ji}, \citenamefont {Wang}, \citenamefont {Qiu},
  \citenamefont {Zhao}, \citenamefont {Ma}, \citenamefont {Huang},
  \citenamefont {Wu},\ and\ \citenamefont {Chang}}]{Yizun:20}%
  \BibitemOpen
  \bibfield  {author} {\bibinfo {author} {\bibfnamefont {Y.}~\bibnamefont
  {He}}, \bibinfo {author} {\bibfnamefont {L.}~\bibnamefont {Ji}}, \bibinfo
  {author} {\bibfnamefont {Y.}~\bibnamefont {Wang}}, \bibinfo {author}
  {\bibfnamefont {L.}~\bibnamefont {Qiu}}, \bibinfo {author} {\bibfnamefont
  {J.}~\bibnamefont {Zhao}}, \bibinfo {author} {\bibfnamefont {Y.}~\bibnamefont
  {Ma}}, \bibinfo {author} {\bibfnamefont {X.}~\bibnamefont {Huang}}, \bibinfo
  {author} {\bibfnamefont {S.}~\bibnamefont {Wu}},\ and\ \bibinfo {author}
  {\bibfnamefont {D.~E.}\ \bibnamefont {Chang}},\ }\bibfield  {title} {\bibinfo
  {title} {Geometric control of collective spontaneous emission},\ }\href
  {https://doi.org/10.1103/PhysRevLett.125.213602} {\bibfield  {journal}
  {\bibinfo  {journal} {Phys. Rev. Lett.}\ }\textbf {\bibinfo {volume} {125}},\
  \bibinfo {pages} {213602} (\bibinfo {year} {2020}{\natexlab{b}})}\BibitemShut
  {NoStop}%
\bibitem [{Note1()}]{Note1}%
  \BibitemOpen
  \bibinfo {note} {{\protect \color {brickred}A more accurate estimate of the
  population dynamics $P_{\protect \mathrm {qu},N_{c}}^{\protect \mathrm
  {col}}(t)$ can be obtained} by considering the effective coupling
  $g_{\protect \mathrm {eff},c}$ as defined in Eq.~\protect \eqref {Eq:Hpsi}.
  However, the variation in individual coupling strengths prevents an explicit
  expression of the Dicke enhancement as $\protect \sqrt {N_{c}}$. Using the
  average $\protect \bar {g}_{c}$ provides a reliable estimate for the expected
  behavior.}\BibitemShut {Stop}%
\bibitem [{\citenamefont {Diniz}\ \emph {et~al.}(2025)\citenamefont {Diniz},
  \citenamefont {Villas-Boas},\ and\ \citenamefont {Santos}}]{Diniz:25}%
  \BibitemOpen
  \bibfield  {author} {\bibinfo {author} {\bibfnamefont {C.~M.}\ \bibnamefont
  {Diniz}}, \bibinfo {author} {\bibfnamefont {C.~J.}\ \bibnamefont
  {Villas-Boas}},\ and\ \bibinfo {author} {\bibfnamefont {A.~C.}\ \bibnamefont
  {Santos}},\ }\bibfield  {title} {\bibinfo {title} {Scalable quantum eraser
  with superconducting integrated circuits},\ }\href
  {https://doi.org/10.1088/2058-9565/adbded} {\bibfield  {journal} {\bibinfo
  {journal} {Quantum Science and Technology}\ }\textbf {\bibinfo {volume}
  {10}},\ \bibinfo {pages} {025039} (\bibinfo {year} {2025})}\BibitemShut
  {NoStop}%
\bibitem [{\citenamefont {Rastogi}\ \emph {et~al.}(2022)\citenamefont
  {Rastogi}, \citenamefont {Saglamyurek}, \citenamefont {Hrushevskyi},\ and\
  \citenamefont {LeBlanc}}]{Rastogi:22}%
  \BibitemOpen
  \bibfield  {author} {\bibinfo {author} {\bibfnamefont {A.}~\bibnamefont
  {Rastogi}}, \bibinfo {author} {\bibfnamefont {E.}~\bibnamefont
  {Saglamyurek}}, \bibinfo {author} {\bibfnamefont {T.}~\bibnamefont
  {Hrushevskyi}},\ and\ \bibinfo {author} {\bibfnamefont {L.~J.}\ \bibnamefont
  {LeBlanc}},\ }\bibfield  {title} {\bibinfo {title} {Superradiance-mediated
  photon storage for broadband quantum memory},\ }\href
  {https://doi.org/10.1103/PhysRevLett.129.120502} {\bibfield  {journal}
  {\bibinfo  {journal} {Phys. Rev. Lett.}\ }\textbf {\bibinfo {volume} {129}},\
  \bibinfo {pages} {120502} (\bibinfo {year} {2022})}\BibitemShut {NoStop}%
\end{thebibliography}
%

\end{document}


\title{Supplementary Material for: \\ Collective Dynamics in Circuit Quantum Acoustodynamics with a Macroscopic Resonator}
	
\author{Libo Zhang}
\thanks{These authors contributed equally to this work.}
\affiliation{\szkl}\affiliation{\gdpkl}

\author{Chilong Liu}
\thanks{These authors contributed equally to this work.}
\affiliation{\szkl}\affiliation{\gdpkl}

\author{Guixu Xie}
\thanks{These authors contributed equally to this work.}
\affiliation{\szkl}\affiliation{\gdpkl}

\author{Haolan Yuan}
\thanks{These authors contributed equally to this work.}
\affiliation{\szkl}\affiliation{\gdpkl}

\author{Mingze Liu}
\affiliation{\szkl}\affiliation{\gdpkl}

\author{Hao Jia}
\affiliation{\szkl}

\author{Jian Li}
\affiliation{\szkl}

\author{Chang-Kang Hu}
\email{huchangkang@iqasz.cn}
\affiliation{\szkl}

\author{Song Liu}
\email{lius@iqasz.cn}
\affiliation{\szkl}\affiliation{\HFNL}

\author{Alan C. Santos~\orcidlink{0000-0002-6989-7958}}
\email{ac\_santos@iff.csic.es}
\affiliation{\CSIC}

\author{Dian Tan}
\email{tandian@iqasz.cn}
\affiliation{\szkl}\affiliation{\HFNL}

\author{Dapeng Yu}
\affiliation{\szkl}\affiliation{\HFNL}
	
	\maketitle
	
	
	\setcounter{equation}{0}
	\setcounter{figure}{0}
	\setcounter{table}{0}

	\renewcommand{\thesection}{S\Roman{section}}
	
	\renewcommand{\theequation}{S\arabic{equation}}
	\renewcommand{\thefigure}{S\arabic{figure}}
	\renewcommand{\bibnumfmt}[1]{[S#1]}
	\renewcommand{\citenumfont}[1]{S#1}

\tableofcontents


    \section{Theory Model}
\label{apendice-prova-MSfases}

In this section we introduce the theoretical model used to describe the properties of our device. As discussed in the main text, we introduce the clustered HBARs as a system described by a set of quantum harmonic oscillators with frequencies $\omega_{n,c}$, and those clusters are almost identically separated in frequency $\Delta_{c} = \omega_{n,c+1} - \omega_{n,c}$~\cite{Crump:23,Chu:17,Chu:18}. In general, these HBAR modes are coupled to qubits with strength $g_{n,c}$ much smaller than detuning between the modes $\Delta_{c}$. In particular, this is the case of single-mode clusters that allow for engineering of effective interaction between different modes~\cite{von_Lupke:24}.

To describe the coupling between the artificial atom and phononic modes of our clustered HBAR in the strong coupling regime, 
we have the following Hamiltonian  which is described by the JC model for the $c$-th cluster as
\begin{align}
	\hat{H}_{c} &= \omega_\mathrm{qu}\hat{a}^{\dagger}\hat{a} + \frac{\alpha}{2}\hat{a}^{\dagger}\hat{a}^{\dagger}\hat{a}\hat{a} +  \sum_{n=1}^{N_{c}}\omega_{n,c} \hat{f}_{n,c}^{\dagger}\hat{f}_{n,c} 
    \nonumber \\
    &+  \sum_{n=1}^{N_{c}} g_{n,c} \left(\hat{f}_{n,c}^{\dagger}\hat{a} + \hat{f}_{n,c}\hat{a}^{\dagger}\right)
    , \label{App:Eq:FullHamiltonian}
\end{align}
where the atom with anharmonicity $\alpha$ interacts with the clustered mode composed of $N_{c}$ discrete bosonic modes, with frequency $\omega_{n,c}$. By moving to the interaction picture, we get
\begin{align}
	\hat{H}_{c}^{\mathrm{int}}(t) &= \frac{\alpha}{2}\hat{a}^{\dagger}\hat{a}^{\dagger}\hat{a}\hat{a} + \sum_{n=1}^{N_{c}} g_{n,c} \left(\hat{f}_{n,c}^{\dagger}\hat{a}e^{i\Delta_{n,c}t} + \hat{f}_{n,c}\hat{a}^{\dagger}e^{-i \Delta_{n,c}t}\right), \label{Eq:H_int}
\end{align}
where $\Delta_{n,c} = \omega_{n,c} - \omega_\mathrm{qu}$ is the detuning between the emitter and the $n$-th mode belonging to the $c$-th cluster.

\subsection{Collective dynamics}

Because the modes are not identical, even when we set the emitter to the frequency of the $n$-th mode, we will observe time-dependent oscillations in the interaction due to the $(n-1)$-th and $(n+1)$-th modes. \textit{It means that static dark and bright states cannot be reached in this system, what makes it a good candidate to a universal qubit resetting device, even if we couple multiple qubits to the HBAR, as done in Ref.~\cite{Crump:23}}.

In fact, the collective behavior of the system can be witnessed by the fact that, by defining the collective mode operators
\begin{align}
	\hat{F}_{c}(t) &= \sum_{n=1}^{N_{n}} g_{n,c} \hat{f}_{n,c}e^{-i\Delta_{n,c}t} ,
\end{align}
the interaction Hamiltonian becomes 
\begin{align}
	\hat{H}_{c}^{\mathrm{int}}(t) &=  \hat{F}_{c}^{\dagger}(t)\hat{\sigma}^{-} + \hat{F}_{c}(t)\hat{\sigma}^{+}, \label{App:Eq:IntH_qubit}
\end{align}
where in this last equation we truncate the artificial atom as a two-level system, which can be done by doing the substitution $\hat{a} \rightarrow \hat{\sigma}^{-}$ in Eq.~\eqref{Eq:H_int}, and using that $\hat{\sigma}^{-}\hat{\sigma}^{-} = 0$. From this equation, it is evident the collective aspect of the interaction, as the atom only can send excitation to the phonon modes through the collective mode $\hat{F}_{c}^{\dagger}(t)$. In addition, it is possible to show that, when applied to the mode ground state $\ket{\emptyset}$, $\hat{F}_{c}^{\dagger}(t)$ gives,
\begin{align}
	\hat{F}_{c}^{\dagger}(t)\ket{\emptyset} =  \sum_{n=1}^{N_{c}} g_{n,c} e^{i\Delta_{n,c}t} \ket*{1_{\omega_{n},c}} ,
\end{align}
where $\ket*{1_{\omega_{n},c}}$ denotes the single-phonon state with frequency $\omega_{n}$ inside the $c$-th mode. This equation is relevant here because the symmetries of the system does not allow for high-phonon generation if we start in a state with a single excitation in the system. In this way, the state of interest in our analysis is a timed Dicke state (or W-state in this case) of the form	
\begin{align}
	\ket{\Dcal_{c}(t)} = \frac{\hat{F}_{c}^{\dagger}(t)\ket{\emptyset}}{\sqrt{G_{c}}} = \sum_{n=1}^{N_{c}} \frac{g_{n,c}e^{i\Delta_{n,c}t}}{\sqrt{G_{c}}}  \ket*{1_{\omega_{n},c}} ,
\end{align}
with $G_{c} = \sum_{n=1}^{N_{c}}g_{n,c}^2$ the normalization function. So, now, notice that this state allows us to write the hybrid HBAR-atom state $\ket{\emptyset}\ket{e}$, with $\ket{e}$ the excited state of the atom. This state provides
\begin{align}
	\hat{H}_{c}^{\mathrm{int}}(t)\ket{\emptyset}\ket{e} &=  \left(\hat{F}_{c}^{\dagger}(t)\ket{\emptyset}\right)\left(\hat{\sigma}^{-}\ket{e} \right)
	= \sqrt{G_{c}}\ket{\Dcal_{c}(t)}\ket{g}
	\nonumber \\
	&= g_{\mathrm{eff},c}\ket{\Dcal_{c}(t)}\ket{g}
	,
\end{align}
what allows us to write the effective coupling as $g_{\mathrm{eff},c} = \sqrt{G_{c}}$. It means that, a perfect dark state can only appear if there exist phases $\phi_{n,c}$ such that $\sum_{n} g^{2}_{n}e^{i\phi_{n}} = 0$. While this condition can be satisfied for static Dicke states, of the form $\ket*{\Dcal_{c,\mathrm{st}}} = \sum_{n} g_{n,c}e^{i\phi_{n,c}}\ket*{1_{\omega_{n},c}}/g_{\mathrm{eff},c}$, for time-independent phases $\phi_{n,c}$, it can be challenging to achieve for dynamically induced timed-Dicke states due to the time-dependent phases $e^{i\Delta_{n,c}t}$. A similar result can be obtained for the super-radiant phase, as we need identical phases $\phi_{n,c} = \phi_{0,c}$, which is not possible because $\Delta_{n,c}$ depend on the frequency of non-degenerate modes.

This allows us then to observe a Dicke and timed-Dicke transition and we can estimate the time interval in which the system approximately behaves as perfect static Dicke states. To this end, we will consider an estimate of such a static-timed-Dicke transition time through the overlap between the static Dicke and timed-Dicke states, $\Fcal_{N_{c}}(t) = |\bra{\Dcal_{c,\mathrm{st}}}\ket{\Dcal_{c}(t)}|$. Without loss of generality, we assume a system intra-cluster frequency modes equally separated by $\delta_{\mathrm{SRF},c} = \omega_{n+1,c}-\omega_{n,c}$, also with identical mode-qubit couplings $g_{n,c}=g_{c}$. Therefore,
\begin{align}
	\Fcal_{N_{c}}(t)  &= |\bra{\Dcal_{c,\mathrm{st}}}\ket{\Dcal_{c}(t)}| = \left \vert\frac{1}{N_{c}} \sum_{n=1}^{N_{c}} e^{i(\omega_{n,c} - \omega_\mathrm{qu})t}\right \vert ^2
	\nonumber\\
	&= \left \vert\frac{1}{N_{c}} e^{i(\omega_{1,c} - \omega_\mathrm{qu})t} \sum_{n=1}^{N} e^{i(\omega_{n,c} - \omega_{1,c})t} \right \vert ^2
	= \left \vert\frac{1}{N_c} \sum_{n=0}^{N_{c}-1} e^{in\delta_{\mathrm{SRF},c} t} \right \vert ^2
	\nonumber\\
	&= \frac{1}{N_{c}^2} \frac{1 - \cos[N_{c} \eta_{c}(t)] }{1 - \cos[\eta_{c}(t)]}, \label{Ap:Eq:Fidel}
\end{align}
where we used that $|e^{i(\omega_{1,c} - \omega_\mathrm{qu})t}|=1$ and $\omega_{n,c} - \omega_{1,c} = n\delta_{\mathrm{SRF},c}$, and where we defined $\eta_{c}(t) = \delta_{\mathrm{SRF},c} t$. From this equation, we can estimate the exact value of $\tau_{\mathrm{timed}}$ required to achieve a target fidelity $\Fcal_{N_{c}}(\tau_{\mathrm{timed}}) = \Fcal_{0}$. 

In particular, we expect that the static Dicke regime is achieved for small values of $\eta_{t}$, as it approximately constitutes a system with degenerate modes. For this reason, we can expand the Eq.~\eqref{Ap:Eq:Fidel} in the limit $\eta_{c}(t) \ll 1$ and we get
\begin{align}
	\Fcal_{N_{c}}(t) &= 1 + \frac{1}{12} \left(1-N_{c}^2\right) \eta_{c}^2(t) + O\left(\eta_{c}^4(t)\right) 
    \nonumber \\
    &\approx 1 + \frac{1}{12} \left(1-N_{c}^2\right) \eta_{c}^2(t) ,
\end{align}
where we can neglect the fourth-order corrections and compute the solution of $\Fcal_{N}(t) = \Fcal_{0}$, which yields the solution
\begin{align}
	\eta_{c}(t) \approx \frac{2 \sqrt{3-3 \mathcal{F}_{0}}}{\sqrt{N^2-1}} ,
\end{align}
as the minimum $\eta_{c}(t)$ to achieve $\Fcal_{N_{c}}(\tau_{\mathrm{timed}}) = \Fcal_{0}$.

\subsection{Qubit purity}

By considering the atom coupled to the clustered multi-mode HBAR, in the single excitation sector the state of the system is described by the Wigner-Weisskopf ansatz
\begin{align}
\ket{\psi_{\mathrm{WW}}(t)} = \left[p(t) \hat{a}^{\dagger} + \sum_{n=1}^{N_{c}}\sum_{c=1}^{N} \vartheta_{\omega_{n},c}(t) \hat{f}^{\dagger}_{\omega_{n},c} \right] \ket{g}\ket{0} ,
\end{align}
where $|p(t)|^2+ \sum_{n=1}^{N_{c}}\sum_{c=1}^{N} |\vartheta_{\omega_{n},c}(t)|^2 = 1$ for wavefunction normalization. With this state, we can then compute the reduced density matrix for the qubit. To this end, we use that $\hat{\rho}_\mathrm{qu}(t) = \mathrm{tr}_{\mathrm{modes}} \left[\hat{\rho}(t) \right]$, where
\begin{align}
\hat{\rho}(t) &= \ket{\psi_{\mathrm{WW}}(t)} \bra{\psi_{\mathrm{WW}}(t)} 
\nonumber \\
%
&=|p(t)|^2 \ket{e}\bra{e} \otimes \ket{0}\bra{0}
\nonumber \\ &+
\sum_{n^{\prime}=1}^{N_{c}}\sum_{c^{\prime}=1}^{N} p^{\ast}(t)\vartheta_{\omega_{n^{\prime}},c^{\prime}}(t) \ket{e} \bra{g} \otimes \ket{0}\bra*{1_{\omega_{n^{\prime}},c^{\prime}}}
\nonumber \\ &+
\sum_{n^{\prime},n=1}^{N_{c^{\prime}},N_{c}}\sum_{c^{\prime},c=1}^{N,N} \vartheta^{\ast}_{\omega_{n},c}(t)\vartheta_{\omega_{n^{\prime}},c^{\prime}}(t) \ket{g}\bra{g}\otimes \ket*{1_{\omega_{n},c}} \bra*{1_{\omega_{n^{\prime}},c^{\prime}}} 
\nonumber \\ &+
\sum_{n=1}^{N_{c}}\sum_{c=1}^{N} \vartheta^{\ast}_{\omega_{n},c}(t)p(t) \ket{g}\bra{e} \otimes \ket*{1_{\omega_{n},c}}\bra{0}  . \label{Ap:Eq:rhostate}
\end{align}

The vector basis for the Hilbert space of the modes with up to one excitation reads as $\{ \ket{0},\ket*{1_{\omega_{n},c}} \}$, therefore we compute the trace over the modes as
\begin{align}
\hat{\rho}_\mathrm{qu}(t) &= \mathrm{tr}_{\mathrm{modes}} \left[\hat{\rho}(t) \right] 
\nonumber \\ &=  \bra{0}\hat{\rho}(t)\ket{0} + \sum_{n^{\prime\prime}=1}^{N_{c^{\prime\prime}}}\sum_{c^{\prime\prime}=1}^{N}\bra*{1_{\omega_{n^{\prime\prime}},c^{\prime\prime}}}\hat{\rho}(t)\ket*{1_{\omega_{n^{\prime\prime}},c^{\prime\prime}}} ,
\end{align}
which gives us
\begin{align}
\hat{\rho}_\mathrm{qu}(t) &= |p(t)|^2 \ket{e}\bra{e} + |p_{g}(t)|^2\ket{g}\bra{g} , \label{Ap:Eq:reducedrho}
\end{align}
where we defined the ground state population for the atom as, given by
\begin{align}
|p_{g}(t)|^2 &= 
 \sum\nolimits_{n=1}^{N_{c}}\sum\nolimits_{c=1}^{N}|\eta_{\omega_{n},c}(t)|^2 .
\end{align}

Now, we can use the wavefunction normalization condition of the Eq.~\eqref{Ap:Eq:rhostate} to get the simplified result $|p_{g}(t)|^2 = 1 - |p(t)|^2$. As expected, this result is compatible with the density matrix normalization condition $\mathrm{tr}(\hat{\rho}_\mathrm{qu}(t)) = |p_{g}(t)|^2 + |p(t)|^2 = 1$ for the Eq.~\eqref{Ap:Eq:reducedrho}. Therefore,
\begin{align}
\hat{\rho}_\mathrm{qu}(t) &= |p(t)|^2 \ket{e}\bra{e} + \left(1 - |p(t)|^2\right)\ket{g}\bra{g} .
\end{align}

To compute the purity, we then obtain $\hat{\rho}_\mathrm{qu}^2(t)$ given by
\begin{align}
\hat{\rho}_\mathrm{qu}^2(t) &= |p(t)|^4 \ket{e}\bra{e} + \left(1 - |p(t)|^2\right)^2\ket{g}\bra{g} ,
\end{align}
which allows us to conclude that
\begin{align}
\Pcal(t) = \mathrm{tr}\left(\hat{\rho}_\mathrm{qu}^2(t)\right) &= |p(t)|^4 + \left(1 - |p(t)|^2\right)^2 \nonumber \\
&=
2|p(t)|^4 + 1 - 2|p(t)|^2 . 
\end{align}

Therefore, by using the solution for the qubit population dynamics discussed in the main text, we get
\begin{align}
\Pcal_{c}(t) &=  2\cos^{4}(g_{\mathrm{eff},c}t) + 1 - 2\cos^{2}(g_{\mathrm{eff},c}t) \nonumber\\ &= \frac{1}{4}(3 + \cos(4g_{\mathrm{eff},c}t)) .
\end{align}

From this result, we can compute the time $\tau_{\mathrm{min}\Pcal}$ required to achieve the maximal entanglement between the qubit and the modes. It is done by minimizing the purity through calculation of the critical points $d\Pcal(t)/dt = 0$. By doing that, we obtain
\begin{align}
\frac{d\Pcal_{c}(t)}{dt} = - 4g_{\mathrm{eff},c}\sin(4g_{\mathrm{eff},c}t) ,
\end{align}
which provides the solution of the condition $d\Pcal(t)/dt = 0$ as $t_n = n \pi / 4g_{\mathrm{eff},c}$. Therefore, the criticality analysis can be concluded from the second derivative test using
\begin{align}
\frac{d^2\Pcal_{c}(t)}{dt^2} = - 16g^2_{\mathrm{eff},c}\cos(4g_{\mathrm{eff},c}t) ,
\end{align}
where the minimum is obtained for all values of $t_n$ in which $d^2\Pcal_{c}(t)/dt^2|_{t=t_n} > 0$. Therefore, it is immediate to see that the shortest time interval $t_n$ such that we have the minimum of the purity is
\begin{align}
\tau_{\mathrm{min}\Pcal} = t_{1} = \frac{\pi}{4g_{\mathrm{eff},c} }.
\end{align}

This estimate is confirmed by the experimental data and discussions presented in the main text.

\section{Experimental Methods}
\subsection{Device fabrication}
\begin{figure*}[htb]
	\includegraphics[width=\linewidth]{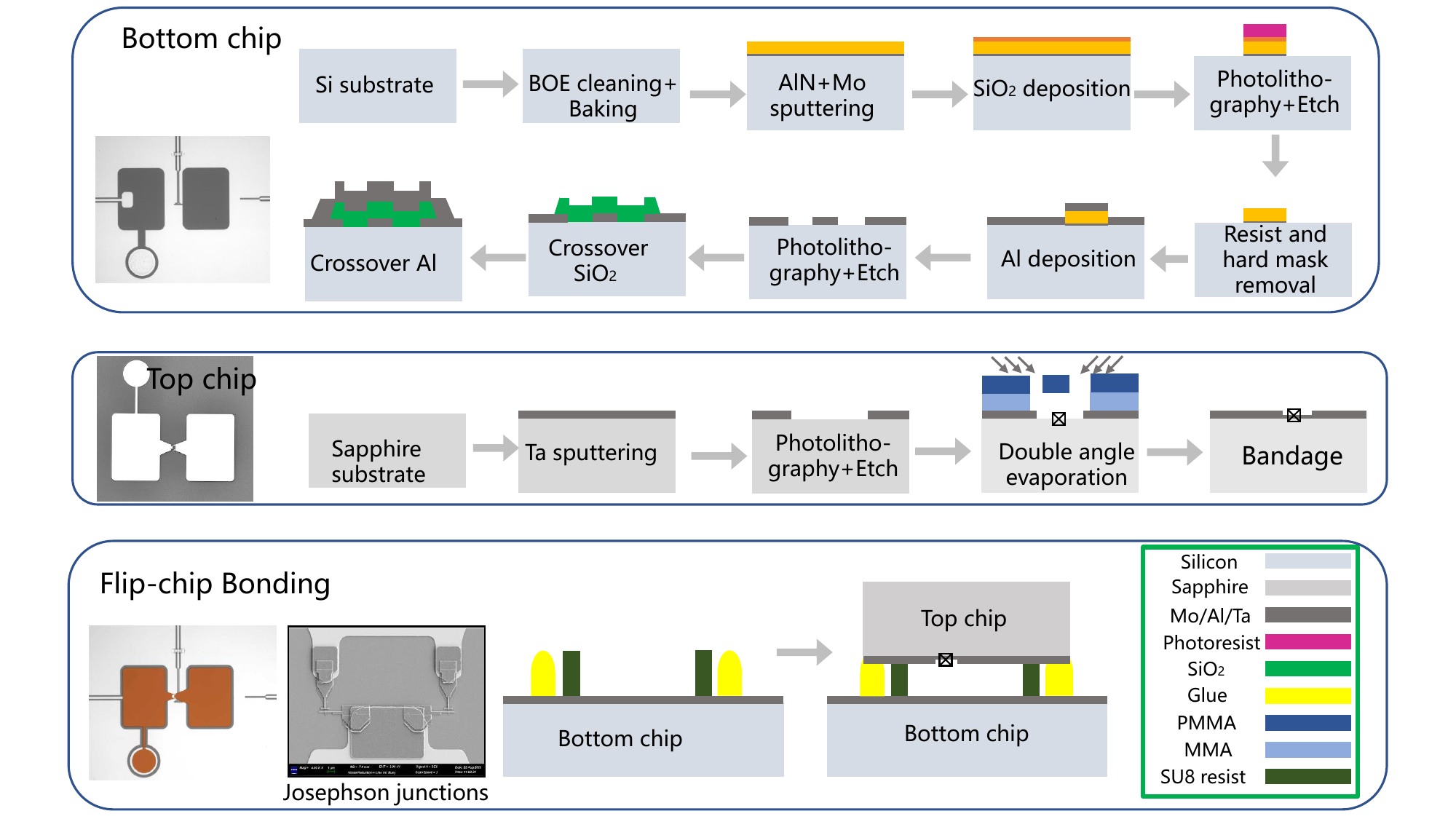}
	\caption{Fabrication of the hybrid quantum system. The bottom chip integrates the AlN/Mo HBAR, control lines, and readout circuitry, fabricated through sputtering, lithography, AlN etching, and Al and SiO$_2$ depositing  processing. The top chip hosts the transmon qubit, formed by Ta patterning, Josephson junction fabrication, and bandage connection. After dicing, the two chips are aligned using SU-8 spacers and bonded by flip-chip assembly to form the final device for hybrid quantum system. }
	\label{Fig_fab}
\end{figure*}
To optimize both the coherence properties of the superconducting qubits and the quality factor of the High-Overtone Bulk Acoustic Resonator (HBAR), we adopt a dual-substrate fabrication strategy in which the qubits and coupling elements are processed on a sapphire top chip, while the HBAR and microwave control/readout circuitry are integrated on a high resistivity silicon bottom chip. After the independent processing of the two substrates, they are aligned and assembled using a flip-chip bonding technique as illustrated in Fig.~\ref{Fig_fab}, enabling strong hybridization between the qubit electric field and the confined acoustic modes of the HBAR.

The fabrication of the top chip begins with the deposition of a 100 nm thick tantalum film onto a c-plane sapphire wafer using DC magnetron sputtering. The growth conditions are optimized to favor the formation of the $\alpha$-phase tantalum, which is known to exhibit low microwave loss and thus supports high-coherence qubit. Following this deposition, the capacitor structures for the qubit  are defined using optical lithography. The patterned regions of the tantalum film are subsequently transferred into the underlying layer by Oxford 100 ICP etching with fluorine-based gases, ensuring well-defined metal edges.

Josephson junctions are fabricated using a standard Dolan-bridge process. Electron-beam lithography is employed to define the small junction area and the undercut required for double-angle evaporation. Aluminum is then deposited at two different angles with an oxidation step in between, forming the Al/AlOx/Al tunnel barrier with the desired critical current in a Plassys evaporation system. A bandage process is used to ensure a reliable galvanic connection between the Josephson junction leads and the neighboring tantalum  capacitor pads. A localized area of argon ion milling removes any native oxide on the tantalum surface, after which an additional layer of aluminum is evaporated to ensure the electrical contact between the junction and the tantalum capacitor.

We first clean the high-resistivity Silicon substrate with BOE solution for 60 seconds and then bake the substrate at 100 degree Celsius for 10 minutes. The bottom chip fabrication then starts with the deposition of a piezoelectric AlN/Mo heterostructure stack on the  silicon substrate in a Clusterline 200 sputtering system (DC Power 7500 W, RF Power 35.5 W, Ar 29 sccm, N$_2$ 87 sccm, Pressure $2.94\times10^{-3}$ mbar). A 900 nm thick layer of highly c-axis-oriented aluminum nitride is deposited by magnetron sputtering, following after a 100 nm  thick of molybdenum layer that serves as both an electrode and an acoustic reflector. A 100 nm thick of SiO$_2$ film is then deposited by chemical vapor deposition to act as a hard mask during the etching of the AlN/Mo films. The HBAR disk region is defined using photo-lithography with Heidelberg direct laser  writer MLA150, after which the exposed SiO$_2$ is etched away by a Sentech ICP with fluorine-based gases to expose the underlying AlN. The AlN/Mo disk layer is obtained through wet etching using $85\%$ concentrated phosphoric acid heated at 100 degree Celsius. After etching, the photoresist is removed with remover 1165 and the remaining oxide mask is etched away using the Sentech ICP.

Once the AlN disk is fabricated, a 100 nm thick  aluminum layer is deposited via a Plassys ebeam evaporator to form the ground plane, microwave electrodes, and the control and readout circuits on the bottom chip. These larger-scale circuit features are defined using optical lithography and wet etching. To allow electrical crossovers between different wiring layers while minimizing microwave crosstalk and slot modes, SiO$_2$ bridges are formed by depositing a 200 nm thick SiO$_2$ film and patterning it through a liftoff process. A 300 nm aluminum layer is then deposited to form the crossover connections after argon ion milling to ensure proper adhesion and electric contact. 


For the flip-chip packaging process, we proceed  following processes. After the above fabrication processes on the two substrates, both chips are diced and cleaned separately. A 2 $\mu$m SU-8 spacer is patterned at the corners of the bottom chip to define the vertical separation between the two chips during assembly. The chips are then aligned with micrometer precision using a flip-chip bonder FC150 , ensuring accurate positioning of the qubit capacitor pads relative to the HBAR electrodes beneath them. A thin layer of nLOF photoresist glue is used to bond the two substrates, forming a mechanically stable stack of the hybrid quantum system that preserves the high-quality of the qubit and the HBAR while enabling strong electromechanical coupling to the superconducting qubit.

\begin{figure}[htbp]
	\centering
	\includegraphics[width=70mm]{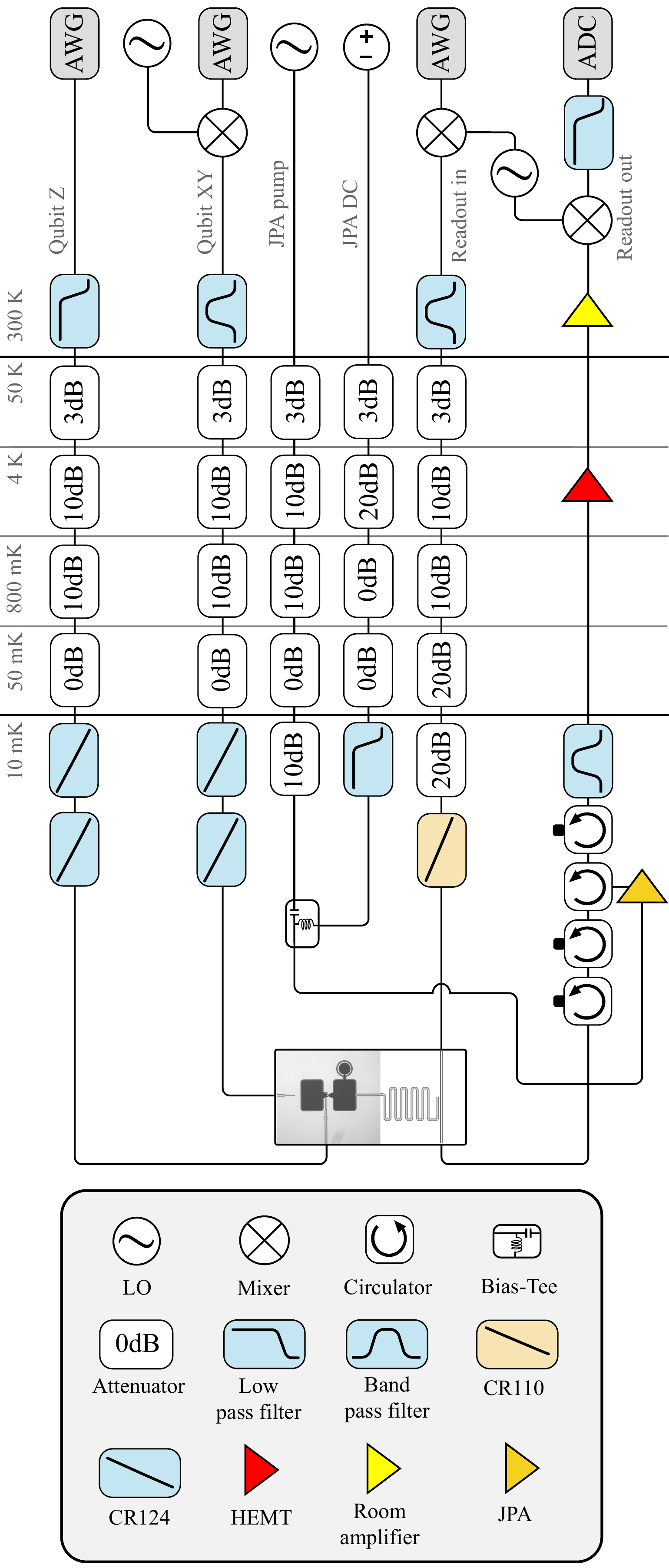}
	\caption{Wiring schematic of the experimental setup.}
	\label{fig: wiring}
\end{figure}

\subsection{Experimental setup}

The experimental setup is depicted in Fig.~\ref{fig: wiring}. The processor was cooled to a base temperature of approximately 10 mK in a dilution refrigerator to minimize thermal noise. Single-qubit control was implemented using programmable microwave pulses resonant with the transition frequency of each qubit. Since arbitrary waveform generators (AWGs) cannot directly output signals in the GHz range, we first generated 600 MHz intermediate-frequency pulses with an AWG. These signals were then up-converted by mixing with a 4.2 GHz local oscillator (LO), followed by bandpass filtering to suppress LO leakage. Z control lines were driven directly by the AWG. The readout signals were generated using an up-conversion scheme similar to that of the qubit control lines. To enhance readout fidelity, a Josephson parametric amplifier (JPA) was employed to provide 15 dB of gain near the readout resonator frequency. The signal was further amplified by a High-Electron-Mobility Transistor (HEMT) at the 4 K stage and additional room-temperature amplifier before down-conversion, sampling with an analog-to-digital converter (ADC), and finally digitized for state discrimination.

\subsection{Readout correction}
\begin{figure}[htbp]
	\centering
	\includegraphics[width=\linewidth]{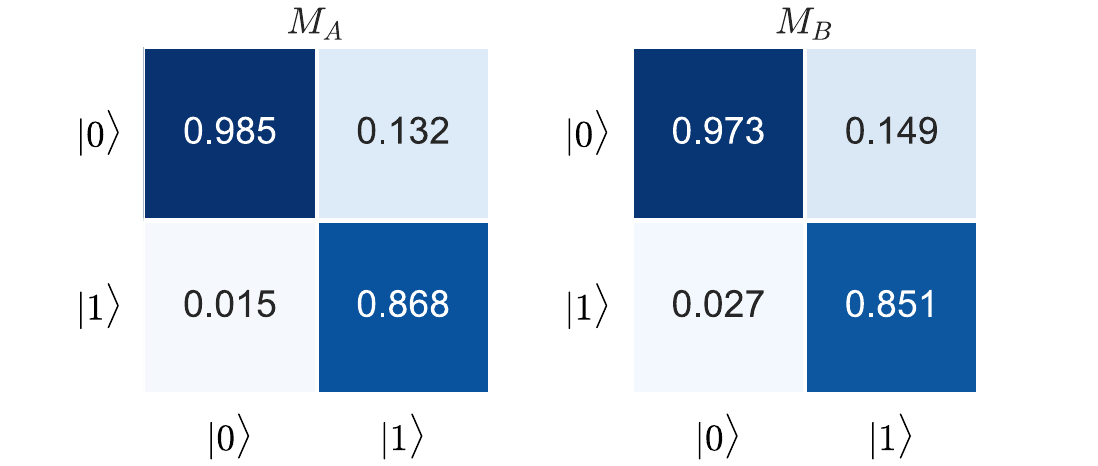}
	\caption{Response matrices for readout correction. The $Q_{A}$ and $Q_{B}$ are two distinct qubits without coupling between them, with each coupled to corresponding HBAR. Their readout corrections are applied separately based on the response matrices.}
	\label{fig: Readoutmatrix}
\end{figure}

The fidelity of experimental outcomes is limited by state preparation and measurement (SPAM) errors as well as intrinsic quantum circuit noise. We can model the measurement errors particularly under conditions of small initialization errors. This model describes the transformation from an ideal probability distribution to a noisy one according to the relation:
\begin{align}\xi_{noisy} = M\xi_{ideal}, ~~ M= \begin{pmatrix}
		p(0|0) & p(0|1) \\
		p(1|0) & p(1|1) 
\end{pmatrix},\end{align}
where $\xi_{ideal}$ and $\xi_{noisy}$ are the vectors of ideal and measured qubit populations, respectively. $M$ is the response matrix. The response matrices for the two qubits, denoted as $M_A$ and $M_B$, are provided in Fig.~\ref{fig: Readoutmatrix}. In our experiment, we consider single qubit readout correction case. Using this model, the ideal measurement result can be simply obtained by the following equation: \begin{align} \xi_{ideal}=M^{-1}\xi_{noisy} \end{align} However, $\xi_{noisy}$ may contain non-physical negative values. To ensure physical results, we employ a least-squares optimization method constrained to non-negative solutions.

\begin{figure*}[htbp]
	\centering
	\includegraphics[width=\linewidth]{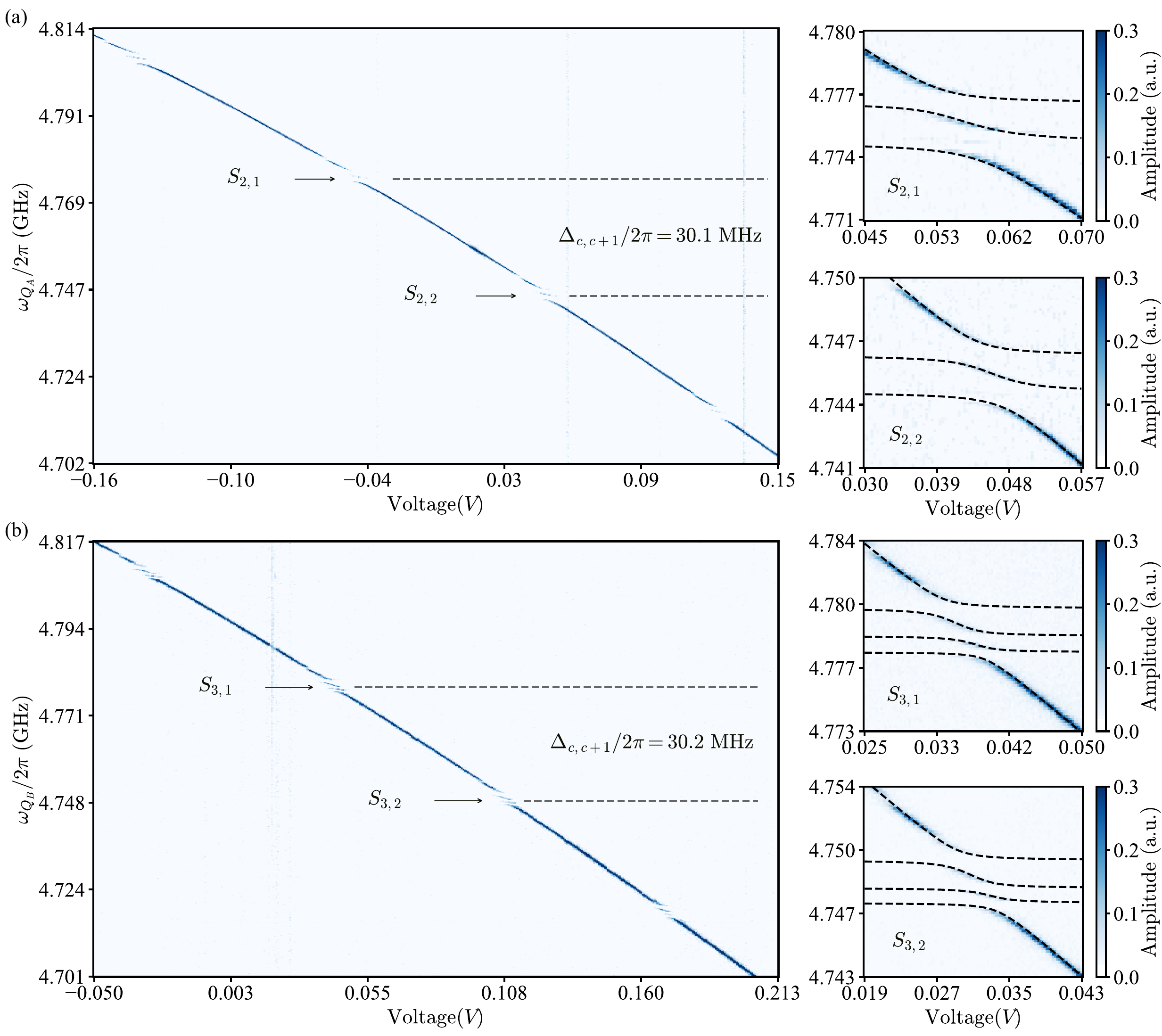}
	\caption{Qubit spectroscopy. (a)-(b) (Left) Anti-crossing between the $Q_A$  ( $Q_B$ ) and the four two-mode (three-mode) HBAR clusters near 4.752 GHz. The $S_{2,1}$ and $S_{2,2}$ ($S_{3,1}$ and $S_{3,2}$) used in the main text are marked. (Right) Fine scan of the anti-crossings near the $S_{2,1}$ and $S_{2,2}$ ($S_{3,1}$ and $S_{3,2}$).}
	\label{fig: anti_crossing}
\end{figure*}

\begin{table}[htbp]
	
	\renewcommand\arraystretch{1}
	\centering
	
	\begin{tabular}{@{} lcc @{}}
		
		\hline
		\hline
		& $Q_A$ & $Q_B$\\
		\hline
		Maximum frequency (GHz)                                 & 4.865 & 4.945 \\
		Minimum frequency (GHz)									& 3.742 & 3.654 \\
		\hline
		Idele frequency (GHz)									& 4.7912 & 4.7938 \\
		Anharmonicity (MHz)                             		& -270  & -289  \\
		Resonator frequency (GHz)                       		& 6.336 & 6.656 \\
		Resonator linewidth (MHz)                       		& 1.03  & 1.90  \\
		Dispersive shift of $|1\rangle$ (MHz)           		& 0.75  & 0.86  \\
		
		Readout fidelity of $|0\rangle$ (\%)           			& 98.5  & 97.3  \\
		Readout fidelity of $|1\rangle$ (\%)            		& 86.8  & 85.1  \\

		Relaxation time of $|1\rangle$  ($\mu s$)       		& 5.13  & 5.51 \\
		Ramsey decay time ($\mu s$)                     		& 2.25  & 1.65 \\
		Spin echo decay time ($\mu s$)                  		& 8.40  & 5.85 \\
		HBAR FSR (MHz) 											& 30.1  & 30.2  \\

		\hline
		\hline
		\\
		\hline
		\hline
		
		& $\omega_c / 2\pi$ (GHz) & $g/2\pi$(MHz)\\
		\hline
		$S_{2,1}$ &4.7766&0.68\\
		& 4.7748 & 0.89\\
		\hline
		$S_{2,2}$ &4.7463&0.70\\
		& 4.7446 & 0.87\\
		\hline
		$S_{3,1}$ &4.7801&0.72\\
		& 4.7785 & 0.58\\
		& 4.7776 & 0.57\\
		\hline
		$S_{3,2}$ &4.7498&0.68\\
		& 4.7481 & 0.62\\
		& 4.7473 & 0.53\\		
		\hline
		\hline

		
	\end{tabular}
	\caption{Device parameters for the hybrid quantum systems. The table presents the measured parameters when the qubits were  at the idle frequency.}
	\label{table:Device}
\end{table}

\subsection{Device parameters} 
Table~\ref{table:Device} summarizes the device parameters for the qubits, readout resonators, and HBARs. The two qubits, denoted as $Q_A$ and $Q_B$, coupling to clustered HBARs of two-mode and three-mode, respectively. The bottom portion of Table ~\ref{table:Device} lists the parameters for the HBAR clusters $S_{2,1}$, $S_{2,2}$, $S_{3,1}$, and $S_{3,2}$, including their frequencies and coupling strengths $g$. The coupling strengths $g$ are determined from the anti-crossings observed in the qubit spectroscopy near these HBAR modes.

\begin{figure*}[htb]
	\centering
	\includegraphics[width=\linewidth]{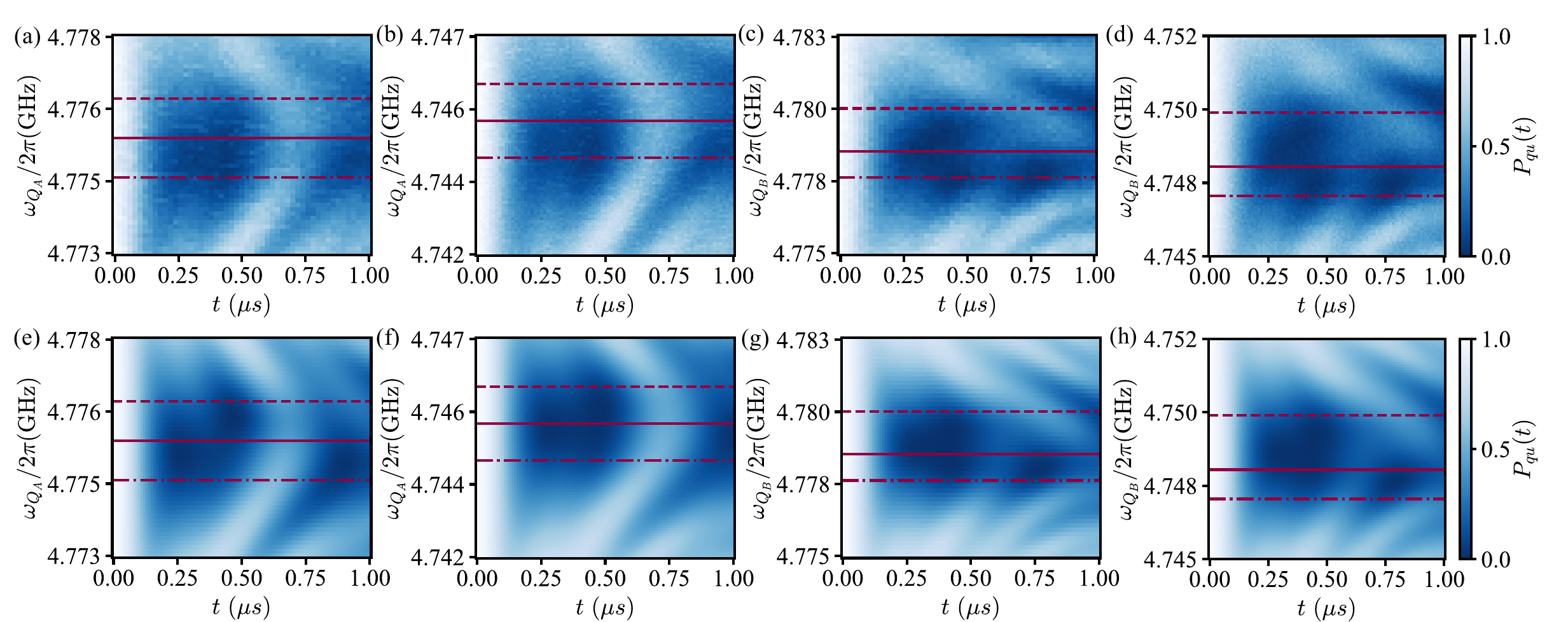}
	\caption{Comparison between the experimental  data and the numerical simulations of the vacuum Rabi oscillations between the qubit and the clusters. (a)-(d) Experimental data of vacuum Rabi oscillations for different HBARs, and (e)-(h) their theoretical counterpart obtained through numerical simulation results for (a)-(d), respectively.}
	\label{fig: Cluster_vac}
\end{figure*}

\subsection{Qubit spectroscopy}In the main text, we presented anti-crossing of $Q_B$ coupled to three-mode and two-mode HBAR clusters. Figure~1, from main text, shows the anti-crossing between Qubit and HBARs from three-mode to two-mode near 4.2 GHz. Due to the relatively short coherence time of the low frequency HBAR modes, we selected higher frequency HBAR modes (around 4.7 GHz) to explore the collective dynamics in the experiments. The HBAR coupled to $Q_A$ exhibits two-mode collective dynamics near 4.7 GHz, whereas the HBAR coupled to $Q_B$ exhibits three-mode collective dynamics in the similar frequencies. The HBAR clusters ($S_{2,1},S_{2,2},S_{3,1},S_{3,2}$) used in the main text is further discussed in detail, where we provide the model and corresponding numerical calculations, which are confirmed by the experimental data.

We now describe the physical system using the full Hamiltonian given in Eq.~\eqref{App:Eq:FullHamiltonian}. Due to the presence of two- or three-mode clusters in HBARs, it is difficult to directly extract the coupling strength $g_{n,c}$ between the qubit and a single mode from the vacuum Rabi oscillations data. Therefore, we perform numerical simulations using the HBAR clusters ($S_{2,1}, S_{2,2},S_{3,1},S_{3,2}$) frequencies and coupling strengths observed from the anti-crossings, which are summarized in Table~\ref{table:Device}. Each coupling strength $g_{n,c}$ is determined by measuring the minimum energy gap ($\Delta f^{\mathrm{min}}_{n,c}$) at the point between the upper and lower frequency branches, where $g_{n,c}/2\pi= \Delta f^{\mathrm{min}}_{n,c}/2$. The center frequency of the interaction corresponds precisely to the frequency at which this minimum gap occurs, representing the resonance point. The simulations are implemented using the QuTiP Python package ~\cite{johansson2012qutip}.

Figure~\ref{fig: anti_crossing} shows the anti-crossings observed in the experimental qubit spectroscopy (left) and the energy eigenvalues of model (right). The qubit frequency can be described as a function of the external voltage (flux). Figure~\ref{fig: anti_crossing}(a) shows the anti-crossing between the Qubit $Q_A$ and HBAR clusters $S_{2,1}$ and $S_{2,2}$ (used in the main text), as well as two nearby HBAR clusters, with FSR of approximately $30.1$ MHz. On the right is a fine scan near HBAR clusters $S_{2,1}$ and $S_{2,2}$. It should be noted that the qubit frequency is set about $15$~MHz above the corresponding HBAR clusters $S_{2,1}$ and $S_{2,2}$ frequencies. The qubit frequency is then tuned downward to HBAR clusters, and the corresponding calculation results based on our Hamiltonian model is shown by the black dashed line. Figure~\ref{fig: anti_crossing}(b) shows the anti-crossing between the Qubit $Q_B$ and HBAR clusters $S_{3,1}$ and $S_{3,2}$, with an FSR of approximately $30.2$~MHz. Similarly, the qubit is set about $15$~MHz above the HBAR clusters frequencies on the right. We find that this hybrid system can be simply modeled by interaction between the qubit and a few dominant modes.

\subsection{Vacuum Rabi oscillations}

The corresponding vacuum Rabi oscillations for the clusters ($S_{2,1}, S_{2,2},S_{3,1},S_{3,2}$) in Figs.~\ref{fig: anti_crossing}(a)-(b) are shown in Figs.~\ref{fig: Cluster_vac}(a)-(d). The system was initialized with the qubit in the excited state and the HBARs in the vacuum state. The system then evolves freely for a variable time~$t$. The state of the qubit is measured to determine its excitation probability. By repeating this sequence for many different evolution times, we obtained the time trace of the qubit excitation probability, which exhibits the characteristic vacuum Rabi oscillations between the qubit and the HBAR clusters.

Due to the inherent collective behavior of our device, addressing the modes individually is challenging. As a consequence, extra effort is required to measure the corresponding phonon lifetime through conventional methods---swap dynamics between the qubit and a single mode in HBARs. To circumvent this problem, we consider an alternative multi-variable approach. Vacuum Rabi oscillations are simulated by varying the qubit frequency and the free evolution time~$t$ from 0 - 1~$\mu$s using the QuTiP Python package ~\cite{johansson2012qutip}. The simulated vacuum Rabi oscillations with the parameters shown in Table~\ref{table:Device} are in good agreement with the experimental data as displayed in Figs.~\ref{fig: Cluster_vac}. 



%
